\documentclass[preprint]{aastex}
\usepackage{multirow}
\newcommand\pcc{\;{\rm cm}^{-3}}
\newcommand\Msun{\; M_{\odot}}
\newcommand\kms{\; {\rm km}\;{\rm s}^{-1}}
\newcommand\ergs{\; {\rm erg}\;{\rm s}^{-1}}
\newcommand\erg{\; {\rm erg}}
\newcommand\mh{\; m_{\rm H}}
\newcommand\cm{\;{\rm cm}}
\newcommand\yr{\; {\rm yr}}
\newcommand\kyr{\;{\rm kyr}}
\newcommand\Myr{\;{\rm Myr}}

\newcommand\pc{\;{\rm pc}}

\newcommand\momunit{\Msun \kms}

\newcommand\Punit{\pcc\,{\rm K}}

\newcommand\Kel{\;{\rm K}}

\newcommand\simgt{\lower.5ex\hbox{$\; \buildrel > \over \sim \;$}}
\newcommand\simlt{\lower.5ex\hbox{$\; \buildrel < \over \sim \;$}}
\newcommand\pderiv[2]{\frac{\partial {#1}}{\partial {#2}}}
\newcommand\deriv[2]{\frac{d {#1}}{d {#2}}}

\newcommand\rbrackets[1]{\left({#1}\right)}
\newcommand\sbrackets[1]{\left[{#1}\right]}

\newcommand\divergence[2][\rbrackets]{\nabla \cdot #1{#2}}

\newcommand\vel{\mathbf{v}}

\newcommand\xhat{\hat{\mathbf{x}} }
\newcommand\yhat{\hat{\mathbf{y}} }

\newcommand\rhat{\hat{\mathbf{r}} }

\newcommand\kbol{k_{\rm B}}
\newcommand\rinit{r_{\rm init}}
\newcommand\Minit{M_{\rm init}}
\newcommand\ESN{E_{\rm SN}}

\newcommand\rfr{r_{\rm free}}
\newcommand\rsnr{r_{\rm snr}}
\newcommand\rhot{r_{\rm hot}}
\newcommand\rst{r_{\rm ST}}
\newcommand\rsf{r_{\rm sf}}
\newcommand\rsfnum{r_{\rm sf}^{\rm n}}
\newcommand\tfr{t_{\rm free}}
\newcommand\tsf{t_{\rm sf}}
\newcommand\tsfc{t_{\rm sf,c}}
\newcommand\tsfw{t_{\rm sf,w}}
\newcommand\tsn{t_{\rm SN}}
\newcommand\tsfnum{t_{\rm sf}^{\rm n}}

\newcommand\tcool{t_{\rm cool}}
\newcommand\vsnr{v_{\rm snr}}
\newcommand\vst{v_{\rm ST}}
\newcommand\vsf{v_{\rm sf}}
\newcommand\Tst{T_{\rm ST}}
\newcommand\Tsf{T_{\rm sf}}
\newcommand\pfr{p_{\rm free}}
\newcommand\pst{p_{\rm ST}}
\newcommand\psf{p_{\rm sf}}
\newcommand\psfnum{p_{\rm sf}^{\rm n}}
\newcommand\psnr{p_{\rm snr}}
\newcommand\Psf{P_{\rm sf}}
\newcommand\ppds{p_{\rm PDS}}
\newcommand\vej{v_{\rm ej}}
\newcommand\Mej{M_{\rm ej}}
\newcommand\Msw{M_{\rm sw}}
\newcommand\Msf{M_{\rm sf}}
\newcommand\Msfnum{M_{\rm sf}^{\rm n}}
\newcommand\Mhot{M_{\rm hot}}
\newcommand\Msh{M_{\rm sh}}
\newcommand\Phot{P_{\rm hot}}
\newcommand\Pamb{P_0}
\newcommand\rhoamb{\rho_0}

\shorttitle{Supernova Momentum Injection}
\shortauthors{Kim and Ostriker}

\begin{document}

\title{Momentum Injection by Supernovae in the Interstellar Medium}

\author{Chang-Goo Kim and Eve C. Ostriker}

\affil{Department of Astrophysical Sciences, Princeton University, Princeton, NJ 08544, USA}
\email{cgkim@astro.princeton.edu, eco@astro.princeton.edu}

\begin{abstract}
Supernova (SN) explosions deposit prodigious energy and momentum in their
environments, with the former regulating multiphase thermal structure and the
latter regulating turbulence and star formation rates in the interstellar
medium (ISM).  However, systematic studies quantifying the impact of SNe in
realistic inhomogeneous ISM conditions have been lacking.  Using
three-dimensional hydrodynamic simulations, we investigate the dependence of
radial momentum injection on both physical conditions (considering a range of
mean density $n_0=0.1-100\pcc$) and numerical parameters.  Our inhomogeneous
simulations adopt two-phase background states that result from thermal
instability in atomic gas.  Although the SNR morphology becomes highly complex
for inhomogeneous backgrounds, the radial momentum injection is remarkably
insensitive to environmental details.  For our two-phase simulations, the final
momentum produced by a single SN is given by $2.8 \times 10^5\momunit
n_0^{-0.17}$. This is only 5\% less than the momentum injection for a
homogeneous environment with the same mean density, and only 30\% greater than
the momentum at the time of shell formation.  The maximum mass in hot gas is
also quite insensitive to environmental inhomogeneity.  Strong magnetic fields
alter the hot gas mass at very late times, but the momentum injection remains
the same.  Initial experiments with multiple spatially-correlated SNe show a
momentum per event nearly as large as single-SN cases.  We also present a full
numerical parameter study to assess convergence requirements.  For convergence
in the momentum and other quantities, we find that the numerical resolution
$\Delta$ and the initial size of the SNR $\rinit$ must satisfy $\Delta,
\rinit<\rsf/3$, where the shell formation radius is given by $\rsf = 30 \pc\;
n_0^{-0.46}$ for two-phase models (or 30\% smaller for a homogeneous medium).
\end{abstract}

\keywords{methods:numerical -- supernovae: general -- ISM: supernova remnants
-- ISM: kinematics and dynamics}

\section{INTRODUCTION}\label{sec:intro}

The importance of supernova (SN) blastwaves to the interstellar medium (ISM)
has been appreciated for many decades
\citep{1974ApJ...189L.105C,1977ApJ...218..148M}.  In recent years, it has
become increasingly clear that accurate treatment of supernova remnant (SNR)
evolution and other energy inputs from massive stars is crucial not just for
detailed models of the ISM and star formation, but also for models of galaxy
formation and evolution. SNe are believed to play the predominant role in
creating the hot phase and shaping the large-scale structure of the multiphase
ISM
\citep[e.g.,][]{1989ApJ...345..372N,2004A&A...425..899D,2012ApJ...750..104H,2014A&A...570A..81H},
in driving turbulence in diffuse gas
\citep[e.g.,][]{1995ApJ...440..634R,2004RvMP...76..125M,2006ApJ...653.1266J},
and in regulating star formation rates by maintaining the large-scale turbulent
pressure that limits gravitational condensation
\citep[e.g.,][]{2011ApJ...731...41O,2011ApJ...743...25K,2013ApJ...776....1K,2012ApJ...754....2S}.
Particularly important, as we shall discuss below, is the momentum injected to
the ISM over the lifetime of a SNR.  The role of SNe in dense, star-forming
molecular clouds is less clear; due to the time delay before SNe occur, other
forms of star formation ``feedback'' including protostellar outflows, the
pressure of ionized gas in \ion{H}{2} regions, forces from direct and
dust-reprocessed radiation, and the pressure of shocked stellar winds may be of
comparable or greater importance in driving turbulence and destroying massive
molecular clouds \citep[see][and references
therein]{2014arXiv1401.2473K,2013arXiv1312.3223D}.  Through their roles in
controlling star formation and launching galactic winds, SNe also regulate the
internal structure of galaxies and the cosmic evolution of galactic
populations, and recent numerical simulations have followed this with
increasingly detailed treatments \citep[e.g.,][]{2003MNRAS.339..289S,
2006MNRAS.373.1074S,2011ApJ...742...76G,
2013ApJ...770...25A,2013MNRAS.434.3142A,
2012MNRAS.421.3522H,2014MNRAS.445..581H}.

The expansion of SNRs in a uniform medium has been extensively studied via
spherically symmetric models and is well characterized by several familiar
stages \citep[e.g.,][]{1972ARA&A..10..129W,2011piim.book.....D}. The SN
explosion event ejects material into interstellar space with typical
kinetic energy of $\sim 10^{51}\erg$ and mass of
$\sim1$-$10\Msun$. The ejecta expand freely as long as the mass swept
up by the forward shock is smaller than the ejecta mass (the free
expansion stage). As the reverse shock heats up the interior, the hot
gas temperature and pressure become very high, and expansion into the
ambient medium proceeds with negligible radiative cooling. This is the
well known phase of evolution analyzed by \citet{1959sdmm.book.....S}
and \citet{1950RSPSA.201..159T} (the ST stage).  In the ST stage, the
outer shock radius varies as $\rsnr\propto t^{2/5}$, and self-similar
solutions describe the interior structure very well. As the temperature drops,
radiative cooling becomes important, and a thin and dense shell is formed at
the outer edge of the SNR where the temperature is lowest and the density is
highest. After shell formation, a pressure-driven snowplow (PDS) stage occurs
while the interior hot gas has non-negligible
pressure \citep[e.g.,][]{1972ApJ...178..159C,1977ApJ...218..148M}.  After the
interior pressure is exhausted, the shell continues to expand and sweep up ISM
gas as a momentum-conserving snowplow.  One-dimensional, spherical hydrodynamic
simulations of SNRs with radiative cooling confirm these main stages, also
showing that simple treatments of the PDS stage (assuming adiabatic evolution
of the hot interior) are inadequate 
\citep[e.g.,][]{1974ApJ...192..457C,1988ApJ...334..252C,
  1998ApJ...500...95T,1998ApJ...500..342B}.

Although spherical models of SNR evolution in a uniform medium are well
developed, direct application to the ISM is questionable since the ISM in
reality is highly inhomogeneous.  Without SNe, the neutral atomic ISM would
consist of two distinct phases, with dense cold neutral medium (CNM) clouds
surrounded by a diffuse warm neutral medium (WNM), as a result of thermal
instability \citep{1965ApJ...142..531F,1969ApJ...155L.149F,
1995ApJ...443..152W,2003ApJ...587..278W}.  Where the CNM collects into a large
enough cloud that the interior is shielded from UV, it becomes molecular
\citep[see e.g.][and references therein]{2014ApJ...790...10S}, and molecular
clouds themselves are highly inhomogeneous because they are pervaded by highly
supersonic turbulence \citep[e.g.][]{2007ARA&A..45..565M}.  The addition of
SN-driven shocks to the ISM makes its inhomogeneity even more extreme,
producing a third phase of extremely tenuous gas
\citep{1977ApJ...218..148M,2005ARA&A..43..337C}.  Although observational
estimates are difficult, the hot ISM is believed  to fill a fraction $\simlt
30\%$ of the Milky Way disk's volume
\citep[e.g.][]{1998ApJ...503..700F,2001RvMP...73.1031F,2007A&A...463.1227K},
with the majority filled by WNM (and embedded cold clouds).

Since CNM clouds are two orders of magnitude denser than the intercloud WNM but
the cold and warm mass fractions of the atomic ISM are comparable
\citep{2003ApJ...586.1067H}, SNR expansion will be altered compared to
spherical models.  Understanding the effects of cloudy structure on SNR
evolution requires multi-dimensional numerical simulations.  As early efforts,
effects of cloud compression and evaporation have been implicitly considered in
spherical numerical simulations
\citep[e.g.,][]{1981ApJ...247..908C,1987MNRAS.224..701W}.  However, most
multi-dimensional numerical simulations have focused on the detailed evolution
and fate of individual shocked clouds 
\citep[e.g.,][]{1994ApJ...420..213K,1994ApJ...433..757M,2005ApJ...619..327F,2006ApJS..164..477N,2008ApJ...680..336S,2008ApJ...678..274O,2013ApJ...766...45J}
rather than SNR evolution as a whole within the cloudy ISM. 

The dearth of systematic studies of the SNR evolution in the multiphase ISM via
direct numerical simulations has also had a serious impact on larger scale
simulations.  Without a model for blastwave propagation in an inhomogeneous
medium during weakly radiative stages, the momentum injection from SN feedback
is sometimes severely underestimated by considering just the momentum of the
immediate ejecta \citep[e.g.,][]{2013ApJ...770...25A}.  When SN feedback is
modeled by thermal energy injection, insufficient numerical resolution leads to
the so-called ``overcooling-problem'' \citep[e.g.,][]{1992ApJ...391..502K},
which artificially suppresses the dynamical impact of SN feedback to the
surrounding medium.  Although it has long been considered a serious problem in
galaxy formation simulations, firm resolution requirements to avoid overcooling
have not yet been quantified.  Instead, many studies simply adopt a
prescription in which cooling is artificially turned off for a period of time
\citep[e.g.,][]{2000ApJ...545..728T,2006MNRAS.373.1074S, 2012MNRAS.423.1726S}.
However, this delayed-cooling prescription has the opposite problem of
extending the ST phase to unrealistically long times.  The method adopted in
\citet{2006MNRAS.373.1074S}, widely used in subsequent studies
\citep[e.g.,][]{2011ApJ...742...76G,2012MNRAS.424.1275B,2013MNRAS.428..129S},
disables cooling for the merging time of \citet{1977ApJ...218..148M}, which
exceeds the shell formation time by an order of magnitude.  This results in the
substantial overestimate of the momentum injection and hot gas production by
SNe.  
To address some of these issues, \citet{2014MNRAS.442.3013K} have recently
developed a (two-phase) subgrid model to follow feedback effects from
superbubbles, in which the proportions of hot and cold subgrid phases are
controlled by thermal conduction, and both phases cool radiatively.  

For the purposes of building analytic models of the ISM and star formation, and
as inputs to subgrid feedback models in numerical simulations of galaxy
formation, two important quantities are the total momentum injected by a SNR,
and the mass of hot gas that it produces.  The momentum, which is transferred
to the warm and cold gas, determines the amplitude of turbulence in the
volume-filling ISM, and also enters inversely in determining the large-scale
star formation rate \citep[e.g.][]{2011ApJ...731...41O,2011ApJ...743...25K}.
The hot gas produced by SNRs is important because it may be able to escape the 
galaxy's gravitational potential and drive a wind
\citep[e.g.][]{1985Natur.317...44C,2000MNRAS.314..511S}.  Although both
quantities are important, momentum is more easily characterized as it is
conserved at late times, whereas the mass of hot gas steadily declines once
cooling has begun.

In this paper, we investigate the momentum injection to the multiphase ISM by
SNe, using direct three dimensional hydrodynamic simulations with cooling.  We
shall show that the total momentum injection to an inhomogeneous medium is
similar to that in a homogeneous medium, with both comparable to the radial
momentum at the end of the ST stage.  This is an order of magnitude larger than
the initial momentum of the SN ejecta.  We shall also provide firm numerical
conditions for resolving the ST stage (thereby obtaining the correct momentum
injection and early evolution of the hot medium) using standard finite-volume
methods for hydrodynamics on a grid.

The remainder of the paper is organized as follows.  In
Section~\ref{sec:theory}, we first summarize theoretical models of spherical
SNR evolution.  We provide analytic estimates of the shell formation time and
key quantities (size, mass, velocity, momentum, temperature) of the SNR at this
epoch. We then carry out three different sets of simulations, with numerical
prescriptions as described in Section~\ref{sec:method}. We present results from
three-dimensional, hydrodynamic simulations for single SNe in a uniform
background in Section~\ref{sec:single}. We delineate the detailed evolution and
compare numerical results to analytic models in Section~\ref{sec:single_ref},
and provide a systematic parameter study to find numerical convergence
conditions in Section~\ref{sec:single_conv}.  In Section~\ref{sec:single_2p},
we perform a similar systematic study for single SNe in the two-phase ISM.  An
initial exploration of effects of multiple correlated SNe in the two-phase ISM
is presented in Section~\ref{sec:multi_TI}.\footnote{
Although for single SNe, the momentum-conserving stage is
typically reached well before the SNR radius approaches the scale height of the
disk, this is not necessarily true for multiple SNe that drive a superbubble.
If a superbubble breaks out of the disk, then energy and mass will be vented to
the halo rather than injecting momentum to the ISM.  In this work, however, we
do not consider disk stratification and breakout.}
Finally, in Section~\ref{sec:sumndis}, we summarize our results and discuss
their implications.
Here, we also relate our work to other recent studies, and discuss physical
limitations of our models and their potential impact on our conclusions.  In
the Appendix, we present results from magnetohydrodynamic models analogous to
those of Section~\ref{sec:single}, with a range of background magnetic field
strengths.

\section{ANALYTIC THEORIES}\label{sec:theory}

In this section, we briefly review analytic theories for the dynamics of
radiative SNRs. Here we consider the simplest case of spherical expansion, for
SN energy $\ESN$ and ejecta mass $\Mej$ in a uniform medium with density of
$\rhoamb$ (see \citealt{1988RvMP...60....1O} for a more comprehensive review).
In simple analytic models, each evolutionary stage of the SNR expansion can be
approximated by a power law expansion of the SNR radius in time as
$\rsnr\propto t^\eta$. The value of the power law exponent $\eta$ distinguishes
four main stages of the expansion
\citep[e.g.,][]{1988ApJ...334..252C,2011piim.book.....D}: free expansion ($\eta
= 1$), Sedov-Taylor ($\eta = 2/5$), pressure-driven snowplow ($\eta = 2/7$),
and momentum-conserving snowplow ($\eta =1/4$).  At each successive stage, the
exponent decreases as the available power to drive expansion declines.  

In the initial stage of evolution, the mass of the ejecta dominates the mass of
material that has been swept up from the circumstellar medium. Thus, the ejecta
expands ballistically into the circumstellar medium with nearly constant
velocity $\vej = (2\ESN/\Mej)^{1/2}$ so that $\rsnr \propto t$ ($\eta =1$).
The free expansion stage ends when the swept-up mass $\Msw = (4\pi/3)
\rhoamb\rsnr^3$ becomes comparable to the ejecta mass. By equating $\Mej=\Msw$,
we obtain the end of the free expansion stage as
\citep[][]{2011piim.book.....D}
\begin{equation}\label{eq:tfree}
\tfr =\frac{\rfr}{\vej} = 464\yr\; \rbrackets{\frac{\Mej}{3\Msun}}^{5/6} E_{51}^{-1/2} n_0^{-1/2},
\end{equation}
for the SNR radius of
\begin{equation}\label{eq:rfree}
\rfr = \rbrackets{\frac{3\Mej}{4\pi\rhoamb}}^{1/3}=2.75\pc\;\rbrackets{\frac{\Mej}{3\Msun}}^{1/3} n_0^{-1/3}.
\end{equation}
Here, $E_{51} \equiv \ESN/10^{51}\erg$, and $n_0 \equiv n_H/1 \pcc$ where the
hydrogen number density of the ambient medium is $n_H=\rhoamb/(1.4\mh)$ for
10\% Helium abundance.

After the reverse shock heats up the ejecta, the evolution of the SNR for
$t>\tfr$ is well described by evolution of an idealized, point source explosion
since the pressure of the SNR far exceeds the ambient medium pressure.  To a
good approximation, in this stage we can neglect the ejecta mass, energy
losses, and the pressure of the ambient medium, and the solution depends only
on $\ESN$ and $\rhoamb$; this defines the ST stage. Simple dimensional
arguments imply that any length scale in the solution at a given time $t$
should be proportional to $(\ESN t^2/\rhoamb)^{1/5}$.  The internal structure
of the SNR in the ST stage is therefore given by a similarity solution with a
similarity variable $\xi \equiv r/(\ESN t^2/\rhoamb)^{1/5}$. A detailed
solution gives $\xi_0 = 1.15167$ at the shock radius, when the specific heat
ratio is $\gamma=5/3$ .  We thus have the shock radius, shock velocity, and the
immediate postshock temperature during the ST stage given by:
\begin{equation}\label{eq:rst}
\rst = 5.0 \pc \;E_{51}^{1/5}  n_0^{-1/5} t_3^{2/5},
\end{equation}
\begin{equation}\label{eq:vst}
\vst = \frac{2}{5}\frac{\rst}{t} = 1.95\times10^3\kms\; E_{51}^{1/5} 
 n_0^{-1/5} t_3^{-3/5},
\end{equation}
\begin{equation}\label{eq:Tst}
\Tst = \frac{3}{16}\frac{\mu \vst^2}{\kbol}=5.25\times10^7 \Kel\; E_{51}^{2/5} 
n_0^{-2/5} t_3^{-6/5} ,
\end{equation}
where $\mu$ is the mean mass per particles, and $t_3\equiv t/10^3\yr$
\citep[e.g.,][]{2011piim.book.....D}.

When radiative energy losses are no longer negligible, the ST stage ends.
Cooling is strongest immediately behind the shock where the density is the
highest, and a cool dense shell bounding the SNR forms.  The shell formation
time can be defined as the time at which the first shocked gas parcel cools
\citep[e.g.,][]{1982ApJ...253..268C, 1986ApJ...304..771C}. For a volumetric
cooling rate $n_H n_e \Lambda(T)$, the cooling time $t_{\rm cool} \equiv
e/|de/dt|$, where $e$ is the internal energy, of the postshock gas is
\begin{equation}\label{eq:tcool}
\tcool = \frac{2.3}{1.2}\frac{k\Tst}{(\gamma+1)n_0\Lambda(\Tst)}.
\end{equation}
A gas parcel shocked at $t_s$ cools down during $\tcool$ and forms a shell at
$\tsf=t_s+\tcool$. The time for first shell formation can be obtained by
substituting for $T_{\rm ST}$ from equation (\ref{eq:Tst}) in equation
(\ref{eq:tcool}) and then finding the time $t_s$ that minimizes $\tsf$:
\begin{equation}\label{eq:tsf}
\tsf = 4.4\times10^4\yr\; E_{51}^{0.22} n_0^{-0.55}.
\end{equation}
Here, we have used an approximate power-law cooling function $\Lambda(T) =
C(T/10^6\Kel)^{-\alpha}$ with $C=1.1\times10^{-22}\ergs\cm^3$ and $\alpha=0.7$
that gives a fairly good fit for our adopted cooling function in a range of
$10^5\Kel<T<10^{7.5}$ \citep[see][]{2011piim.book.....D}. Note that alternative
approximations for the cooling function with $\alpha=0.5-1.0$ change the
exponents for energy and density only slightly, in the range  0.21 to 0.24 and
-0.57 to -0.53, respectively.  At the time of shell formation, the outer
radius, velocity, postshock temperature, and total swept-up mass are given by:
\begin{equation}\label{eq:rsf}
\rsf = 22.6 \pc\; E_{51}^{0.29} n_0^{-0.42},
\end{equation}
\begin{equation}\label{eq:vsf}
\vsf = 202\kms\; E_{51}^{0.07} n_0^{0.13},
\end{equation}
\begin{equation}\label{eq:Tsf}
\Tsf = 5.67\times10^5 \Kel\; E_{51}^{0.13} n_0^{0.26},
\end{equation}
\begin{equation}\label{eq:Msf}
\Msf = 1680 \Msun\; E_{51}^{0.87} n_0^{-0.26}.
\end{equation}
It is notable that the velocity, post-shock temperature, and total swept-up
mass of the remnant at the time of shell formation are insensitive to the
ambient density.  Physically, this is simply because cooling becomes quite
strong when the temperature falls below $\sim 10^6$K and C and O acquire
electrons that can be collisionally excited
\citep[e.g.][]{2011piim.book.....D}; reaching this temperature (and the
corresponding shock speed) essentially defines the radiative stage of a SNR
\citep{1978ppim.book.....S}.

At shell formation, the shock stalls, and the hot gas in the outer portion of
the SNR is rapidly compressed into a cold thin shell. After shell formation,
the shell is still pushed outward by overpressured hot gas in the interior of
the SNR.  Under certain idealizations, this leads to a so-called
pressure-driven snowplow (PDS) stage. The equation of the motion for the thin
shell in the idealized PDS stage can be written as
\begin{equation}\label{eq:shell}
\deriv{(\Msh\vsnr)}{t} = 4\pi \rsnr^2 (\Phot-\Pamb),
\end{equation}
where $\Msh$ is the mass of the shell and $\vsnr=d\rsnr/dt$ is the expansion
velocity of the SNR.  If the radiative cooling of the hot interior gas is
negligible (i.e. only the gas immediately behind the shock is assumed to cool)
and the mass of hot gas is assumed to be constant, the interior pressure would
drop only from adiabatic expansion as $\Phot\propto \rsnr^{-3\gamma}$.  For
$\gamma=5/3$, setting $\Msh = (4\pi/3)\rsnr^3 \rho_0$ and dropping $\Pamb$ on
the right-hand side leads to a self-similar solution with $\eta = 2/7$ (see
\citealt{1988RvMP...60....1O} for possible variations of this equation with
other terms).  For the idealized PDS solution, the SNR interior is treated as
uniform, with a mean pressure 
\begin{equation}\label{eq:Photpds}
\Phot = P_{\rm hot,sf}\rbrackets{\frac{\rsnr}{\rsf}}^{-5}=
P_{\rm hot,sf}\rbrackets{\frac{t}{\tsf}}^{-10/7}, 
\end{equation}
with 
\begin{equation}\label{eq:Photsf}
P_{\rm hot,sf}=\frac{E_{\rm th,ST}}{2\pi\rsf^3}=2.4\times10^6\kbol\Punit\;E_{51}^{0.13}n_0^{1.26},
\end{equation}
where $E_{\rm th,ST}=0.717E_{\rm SN}$.  Note that the mean pressure is about a
half of the post-shock pressure.

The final classical stage of a SNR begins when the interior pressure $\Phot$
has decreased sufficiently that it no longer exceeds the pressure of the
ambient medium $\Pamb$, making the right hand side of Equation (\ref{eq:shell})
zero. This leads to the constant radial momentum and a self-similar expansion
with $\eta =1/4$. 

We now summarize the momentum injection to the ISM from each of the above SNR
expansion stages.  In the free expansion stage, all the momentum is contained
in the ejecta, and the magnitude of the radial momentum is
\begin{equation}\label{eq:pfree}
\pfr = \Mej \vej = 1.73\times10^4\momunit \; \rbrackets{\frac{\Mej}{3\Msun}}^{1/2} E_{51}^{1/2}.
\end{equation}
Since free expansion by definition assumes negligible interaction with the
circumstellar and/or interstellar medium, the ambient medium acquires no
momentum at this stage. 

In the ST stage, the propagating strong shock heats and accelerates the ambient
medium. Using the ST self-similar solutions, the total radial momentum of the
shocked hot gas can be obtained as
\begin{equation}\label{eq:pst}
\pst = \int \rho v 4\pi r^2 dr = 2.69 \rhoamb \vst \rst^3 = 
2.21\times10^4\momunit\; E_{51}^{4/5} n_0^{1/5} t_3^{3/5}.
\end{equation}
The momentum at the time of shell formation (the end of the ST stage) is
obtained by substituting equation (\ref{eq:tsf}) in equation (\ref{eq:pst}):
\begin{equation} \label{eq:psf}
\psf = 2.17\times10^5 \momunit\; E_{51}^{0.93} n_0^{-0.13}.
\end{equation}
We note that this is nearly linear in the SN energy, and is quite insensitive
to the ambient density.  This is because the momentum of the SNR during the ST
stage is $p\propto E/v$, and the expansion velocity at the time of shell
formation from equation (\ref{eq:vsf}) is very insensitive to both total energy
and ambient density, $\vsf \sim 200 \kms$. 

In the PDS stage, the mass and radial momentum in the SNR are dominated by the
shell. Under the assumption that the interior expands adiabatically (which is
not in fact satisfied; see below), the momentum that would be acquired in an
idealized PDS stage after $\tsf$ can be obtained by integrating Equation
(\ref{eq:shell}), leading to total momentum:
\begin{equation}\label{eq:ppds}
\ppds = \psf\sbrackets{1+0.66\int_1^{t_*} \frac{dt_*}{r_*^3}} 
= \psf\sbrackets{1+4.6(t_*^{1/7}-1)},
\end{equation}
where $t_*\equiv t/\tsf$ and $r_*\equiv \rsnr/\rsf$.  Although equation
(\ref{eq:ppds}) suggests only a very slow growth of momentum with time, the
increase over $\psf$ would in principle be substantial if the interior of the
remnant remained hot for $t\gg \tsf$.

However, the above idealized treatment of the post-shell-formation evolution
does not agree with detailed time-dependent solutions, and momentum injection
in practice increases only modestly subsequent to shell formation.  As we shall
show in the following sections \citep[see also][]{1988ApJ...334..252C}, the
classical PDS assumption of a hot, adiabatic interior with no mass loss does
not apply. In fact, the mass of hot gas interior to the shell steadily
decreases in time, because the outermost part flows into the shell and cools.
Since the pressure decreases more rapidly than the simple adiabatic
expectation, there is only a brief, cooling-modified PDS stage. The momentum
injection after shell formation is quite limited, amounting to only $\sim 50\%$
of $\psf$. 

\section{NUMERICAL METHODS}\label{sec:method}

We solve the hydrodynamics equations with cooling using the {\it Athena} code,
which employs an unsplit Godunov algorithm
\citep{2008ApJS..178..137S,2009NewA...14..139S}. Among the various solvers
provided, we utilize the simple MUSCL-Hancock type predictor-corrector scheme
(van Leer integrator; \citealt{2009NewA...14..139S}). For robustness, we
utilize piecewise linear spatial reconstruction 
and  Roe's Riemann solver
with H-correction
\citep{1998JCoPh.145..511S} and first order flux correction
\citep{2009ApJ...691.1092L}. We apply H-correction only for zones with maximum
signal speed difference ($\eta$-coefficient in \citealt{1998JCoPh.145..511S})
larger than $100\kms$, to avoid artificial diffusion at the interface of the
CNM and WNM where the signal speed difference is about $10\kms$. We solve the
cooling term in an operator split manner using a fully implicit method, with
Newton-Raphson root finding. In addition to the Courant condition, the time
step is also limited by the cooling solver, which allows only a factor of two
variation in the local temperature from the value at the previous time step
\citep{2008ApJ...681.1148K}.

The governing equations are
\begin{equation}\label{eq:cont}
\pderiv{\rho}{t}+\divergence{\rho\vel}=0,
\end{equation}
\begin{equation}\label{eq:mom}
\pderiv{(\rho \vel)}{t}+\divergence{\rho\vel\vel + P} =0,
\end{equation}
and 
\begin{equation}\label{eq:energy}
\pderiv{E}{t}+\divergence{(E+P)\vel}=-\rho\mathcal{L}
\end{equation}
where $\rho$ is the mass density, $\vel$ is the velocity, $E\equiv P/(\gamma-1)
+ \rho v^2/2$ is the total energy density, $P$ is the gas pressure, and
$\gamma$ is the ratio of specific heats.   The gas temperature is
$T=P/(1.1n_H\kbol)$ for neutral gas and $T=P/(2.3 n_H\kbol)$ for fully ionized
gas, where the hydrogen number density is $n_H=\rho/(1.4\mh)$ for $10\%$ of
Helium abundance. Since we did not follow the ionization and recombination in
detail, we calculate the gas temperature using the equation for the neutral gas
so that the temperature in our simulations are higher by factor of $2.3/1.1$
for ionized gas ($T\simgt10^4\Kel$). However, the temperature in practice is
only used to obtain the cooling rate, and only indirectly affects the dynamical
evolution.

The net cooling rate per unit volume is $\rho\mathcal{L}\equiv
n_H[n_H\Lambda(T) - \Gamma]$.  For the cooling function $\Lambda(T)$ at low
($T<10^{4}\Kel$) and high ($T>10^{4}\Kel$) temperature gas, we respectively
adopt fitting formulae from \citet[][see also
\citealt{2008ApJ...681.1148K}]{2002ApJ...564L..97K} and piecewise power-law
fits to the cooling function of \citet{1993ApJS...88..253S} with collisional
ionization equilibrium at solar metallicity.  Heating is only applied at
$T<10^4\Kel$, to model photoelectric heating of the warm/cold ISM.  As the
photoelectric heating rate is proportional to the star formation rate, and the
pressure and star formation rate are approximately proportional to each other
for self-consistent solutions
\citep{2010ApJ...721..975O,2011ApJ...743...25K,2013ApJ...776....1K}, we adopt a
heating rate that varies with mean density as $\Gamma/\Gamma_0=(n_H/2\pcc)$,
where $\Gamma_0 =2\times 10^{-26}\ergs$. We neglect thermal conduction in this
study
(see discussion in Section~\ref{sec:sumndis}).

We have run sets of simulations of three different types: (1) a single SN in a
uniform background medium (SU), (2) a single SN in a two-phase medium (S2P),
and (3) multiple SNe in a two-phase medium (M2P). For all three types of
simulation, we vary the mean density of the background medium $n_0\equiv
n_H/1\pcc$ from 0.1 to 100 (10 for M2P). For our standard models, we fix the
total energy of a single SN to be $\ESN=10^{51}\erg$.  Only for comparison to
M2P models, we also have run a single-explosion model with $\ESN=10^{52}\erg$
(E2P).  Since our main interest is on the momentum injection to the ISM at
later times, we ignore the ejecta and hence the free expansion stage.  We
initialize the SNR in all cases within a sphere of radius $\rinit$ ($\rinit$
varies; see below).  The SNR initialization region consists of all cells whose
centers are at a distance $< \rinit$ from the site of SN explosion; within the
SNR initialization region, we either inject thermal energy at uniform density
(most models), or apply a radially-dependent ST solution.

In order to quantify the evolution, we require definitions for the different
gas components.  We define the hot gas as all zones with  $T>2\times10^4\Kel$,
the shell gas as zones with $T<2\times10^4\Kel$ and $v_r>1\kms$, and the
ambient medium as the remainder (note that ambient gas is initially stationary
for these models).  For the ambient medium, we define the cold and warm neutral
medium (CNM and WNM) with temperature cuts of $T<184\Kel$ and $T>5050\Kel$,
respectively. Hereafter, ``SNR'' refers to both hot and shell gas. We define
the radius of the SNR using the mass-weighted mean radius of the shell gas as
$\rsnr \equiv \sum_{\rm shell} \rho r dV/ \sum_{\rm shell} \rho dV$ where $dV=
\Delta^3$. This is approximately equal to the shock radius before shell
formation, and slightly smaller (due to non-negligible thickness of the shell)
than the shock radius after shell formation (see Figure~\ref{fig:rprof}).  The
kinetic and thermal energies are calculated as $E_{\rm kin}\equiv\sum(1/2)\rho
v^2 dV$ and $E_{\rm th}\equiv \sum [P/(\gamma-1)] dV$ for 
the SNR (hot and shell gas),
and the total energy $E_{\rm tot}\equiv E_{\rm kin} + E_{\rm th}$. 
The total radial momentum of the SNR is calculated by $\psnr\equiv \sum \rho
\vel\cdot\rhat dV$. We also measure the total mass of the shell and hot gas,
$\Msh\equiv\sum_{\rm shell} \rho dV$ and $\Mhot\equiv\sum_{\rm hot} \rho dV$,
respectively, as well as the mean pressure of the hot gas $\Phot\equiv\sum_{\rm
hot} P dV/\sum_{\rm hot} dV$.  

\section{Single SN in Uniform Medium}\label{sec:single}

We first perform a set of numerical experiments with a single SN in a uniform,
unmagnetized medium (SU models).\footnote{We have also performed
simulations with a magnetized medium. We find quite similar evolution and
values of physical quantities overall; small differences are found only for the
case of strong magnetic fields ($\beta=0.1$) at late times. 
See Appendix for details.}
For reference, we list in Table~\ref{tbl:uniform} the theoretical estimates of
the shell formation time $\tsf$ as well as the theoretical SNR radius,
postshock temperature, swept-up mass, and total radial momentum at $\tsf$ (see
Equations (\ref{eq:tsf}), (\ref{eq:rsf}), (\ref{eq:Tsf}), (\ref{eq:Msf}), and
(\ref{eq:psf})), for each choice of ambient density in the range $n_0=$ 0.1 to
100.  As we shall describe in Section \ref{sec:single_conv}, in addition to
varying the physical density of the ambient medium, for our SU models we also
cover a wide range of parameter space for the initial radius of the SNR
$\rinit$ and the grid spacing $\Delta$.   This latter set of tests enables us
to evaluate the dependence of the momentum injection on numerical resolution,
and therefore establish minimum resolution requirements for modeling SN
feedback in complex numerical simulations.  Table~\ref{tbl:uniform} also lists
(in parentheses) the numerically-measured values ($\tsfnum$, $\rsfnum$,
$\Msfnum$, and $\psfnum$) obtained from high-resolution simulations, as
described in Section~\ref{sec:single_ref}.

\begin{deluxetable}{lccccccc}
\tabletypesize{\footnotesize} \tablewidth{0pt} 
\tablecaption{Physical Quantities at Shell Formation for 
Uniform Ambient Density
\label{tbl:uniform}} 
\tablehead{ 
\colhead{Model} &
\colhead{$n_H$} &
\colhead{$\tsf(\tsfnum)$} & 
\colhead{$\rsf(\rsfnum)$} & 
\colhead{$\Tsf$} &
\colhead{$\Msf(\Msfnum)$} & 
\colhead{$\psf(\psfnum)$} \\
\colhead{}&
\colhead{[$\pcc$]}& 
\colhead{[kyr]}&
\colhead{[pc]}&
\colhead{[$10^6\Kel$]}&
\colhead{[$10^3\Msun$]}&
\colhead{[$10^5\momunit$]}
}
\startdata 
SU-n0.1  & 0.1  & 156(150)   & 59.5(58.5) & 0.31 & 3.08(2.96) & 2.94(2.78)\\
SU-n1    & 1    & 43.7(41.9) & 22.6(22.5) & 0.57 & 1.68(1.68) & 2.17(2.05)\\
SU-n10   & 10   & 12.2(10.6) & 8.56(8.34) & 1.04 & 0.92(0.85) & 1.60(1.43)\\
SU-n100  & 100  & 3.43(2.63) & 3.25(3.03) & 1.90 & 0.50(0.41) & 1.18(0.97)
\enddata
\tablecomments{
Column 1: model name.
Column 2: ambient medium hydrogen number density.
Column 3: theoretical estimate of shell formation time $\tsf$ 
(Equation (\ref{eq:tsf})) and numerical measure $\tsfnum$
(in parentheses).
Columns 4-7: theoretical estimates and numerical measures (in parentheses) of
radius, postshock temperature, swept-up mass, and total momentum at 
shell formation.
}
\end{deluxetable}

\subsection{High Resolution Reference Runs}\label{sec:single_ref}

\begin{figure}
\epsscale{1.0}
\plotone{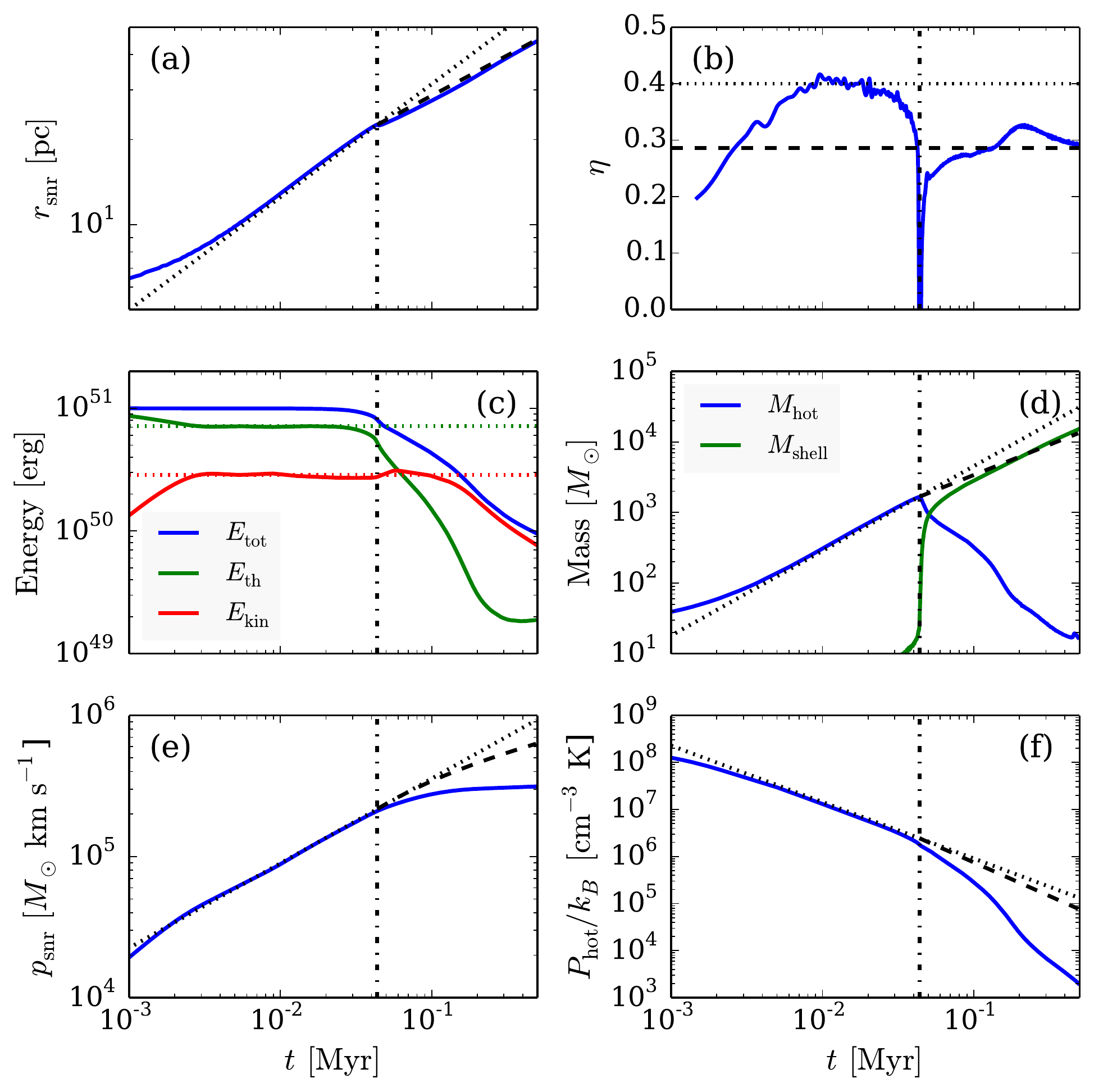}
\caption{
Time evolution of model SU-n1 ($n_0=1\pcc$ and $\ESN=10^{51}\erg$): (a)
mass-weighted radius, (b) deceleration parameter $\eta\equiv \vsnr t/\rsnr$,
(c) total, thermal and kinetic energies, (d) mass of interior hot gas and
shell, (e) total radial momentum, and (f) pressure of interior hot gas. Grid
resolution is $\Delta = 1/4\pc$ and the initial SNR radius is $\rinit = 5\pc$.
The vertical dot-dashed lines in each panel denote the predicted shell
formation time $\tsf=4.4\times10^4\yr$ (Equation (\ref{eq:tsf})) for this model.
The dotted and dashed lines are ST and idealized PDS solutions (at $t>\tsf$)
for each physical quantity (see Section~\ref{sec:theory}).
\label{fig:tevol}}
\end{figure}

\begin{figure}
\epsscale{1.0}
\plotone{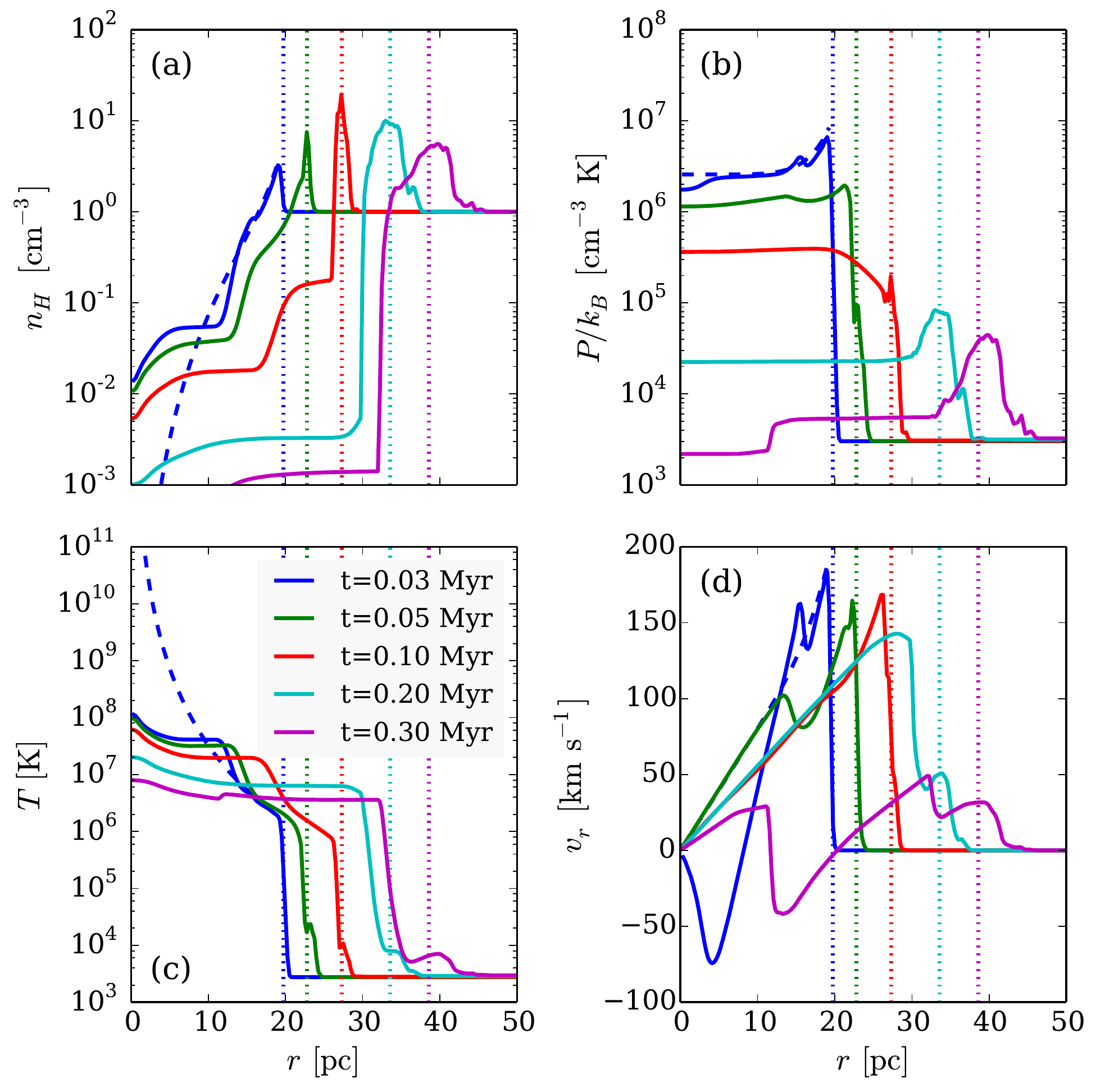}
\caption{
Radial profiles of (a) number density, (b) pressure, (c) temperature, and (d)
radial velocity in model SU-n1 with $\Delta = 1/4\pc$ and $\rinit = 5\pc$.  The
blue dashed line denotes the ST solution at $t=0.03\Myr$.  Vertical dotted
lines are the mass-weighted radius of SNR ($\rsnr$) at each epoch, which agrees
well with the shock position and mean radius of shell before and after shell
formation (at $\tsf=0.04 \Myr$ for this model), respectively.
\label{fig:rprof}}
\end{figure}

\begin{figure}
\epsscale{1.0}
\plotone{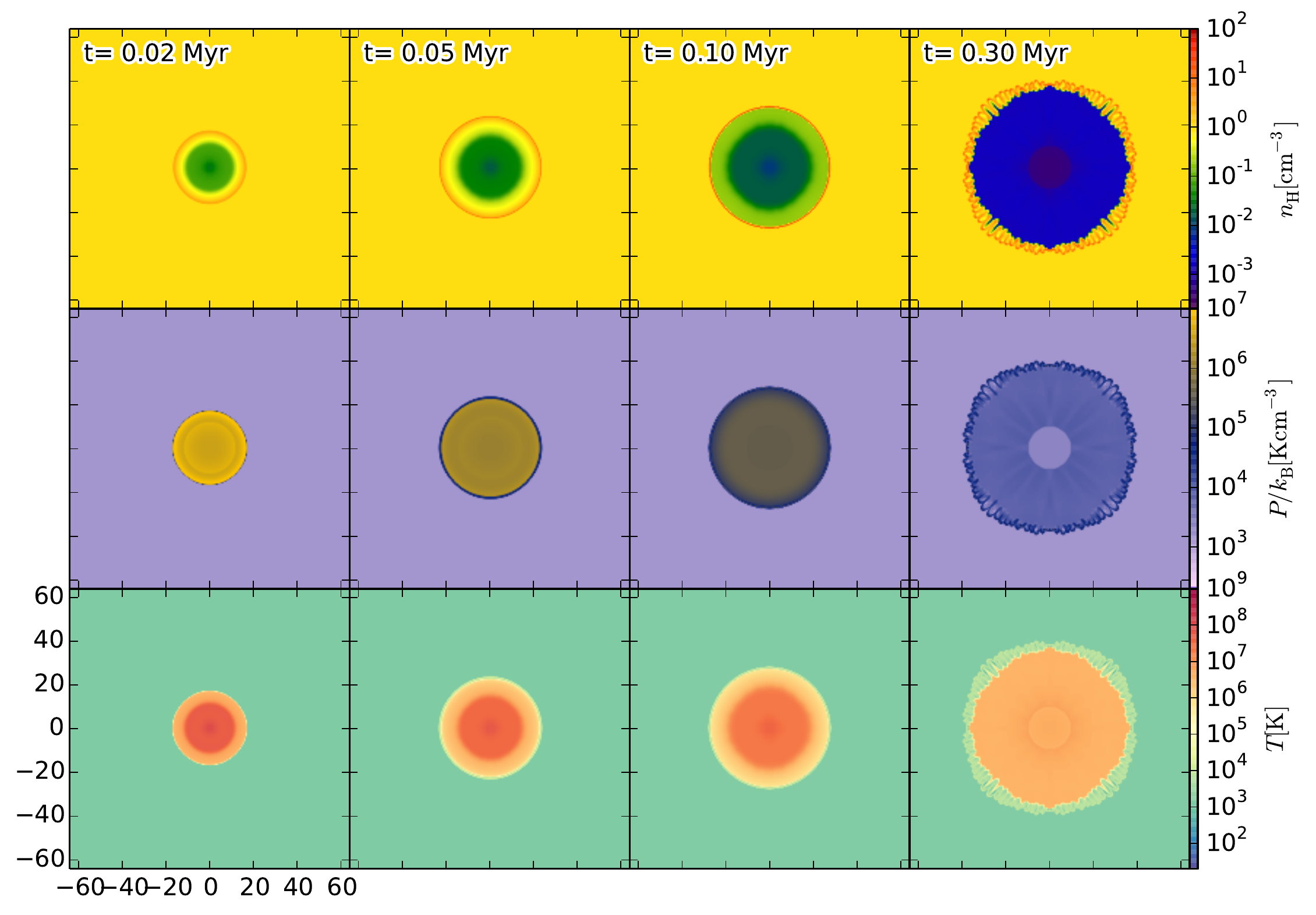}
\caption{
XY-slices from model SU-n1 with $\Delta = 1/4\pc$ and $\rinit = 5\pc$.  From
top to bottom, logarithmic color scales show number density, pressure, and
temperature. From left to right, columns correspond to snapshots at
$t/\tsf\sim1/2$, 1, 2, and 7. A thin shell bounding the SNR is evident in the
two middle columns; in the right column it has become corrugated due to 
dynamical instabilities.
\label{fig:snap}}
\end{figure}

\begin{figure}
\epsscale{1.0}
\plotone{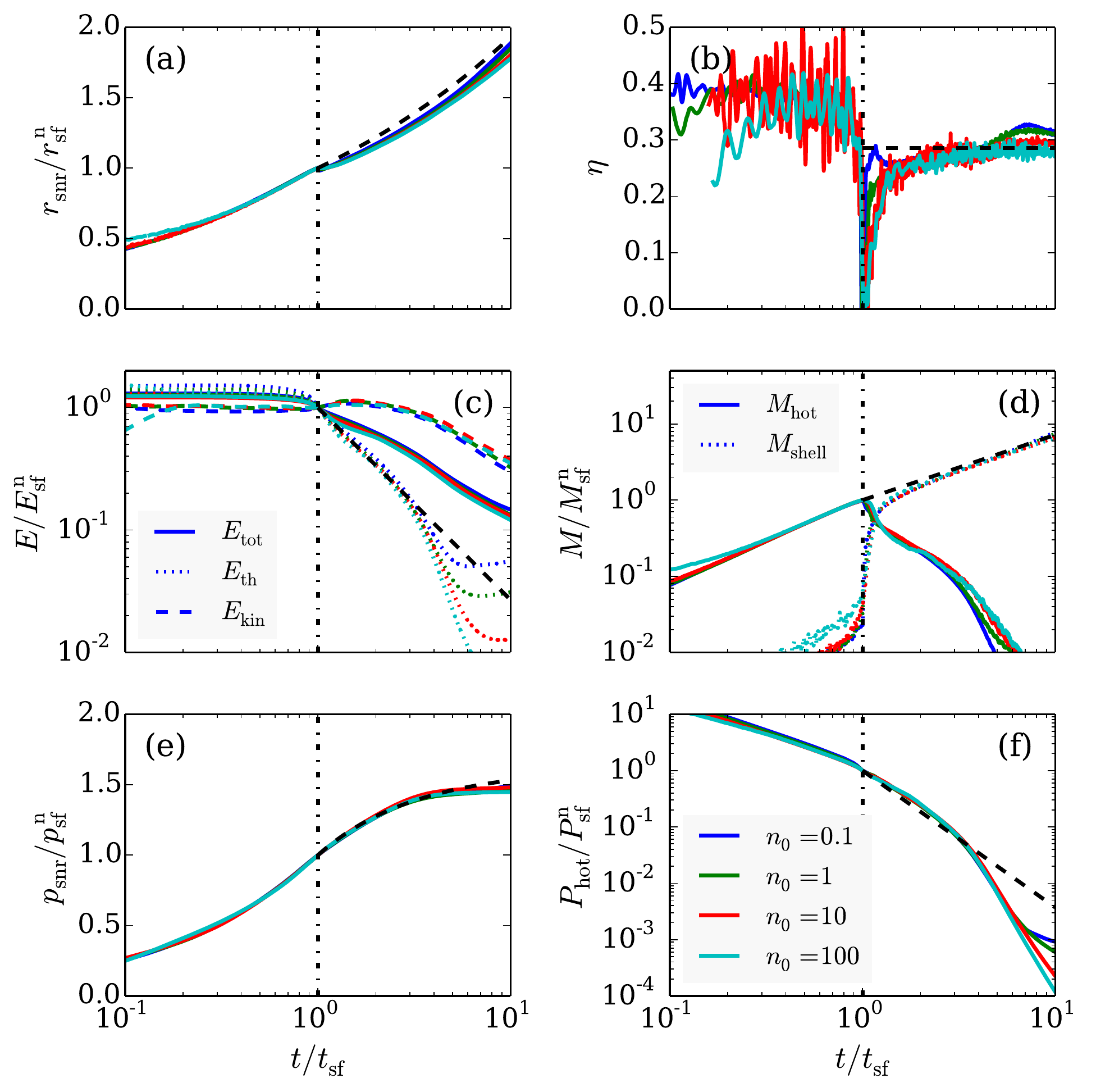}
\caption{
Time evolution comparison for all of the SU high-resolution models, with
ambient medium density $n_0=0.1$ to $n_0=100$.  Panels show (a) mass-weighted
radius, (b) deceleration parameter $\eta\equiv \vsnr t/\rsnr$, (c) total,
thermal and kinetic energies, (d) mass of hot gas and mass of the shell, (e)
total radial momentum, and (f) pressure of interior hot gas.  All physical
quantities are normalized by the corresponding numerical measures at
$t=\tsfnum$ (see text).  The black dashed curves are cooling-modified PDS
solutions after shell formation (see text for details).
\label{fig:tevol_norm}}
\end{figure}

In this subsection, we delineate the time evolution of the SNR based on high
resolution simulations of a single SN in a uniform ambient medium (SU models).
We first present results of SU-n1 with grid resolution $\Delta = 1/4\pc$. The
initial conditions are realized by uniformly applying the SN's thermal energy
($10^{51}\erg$) within an initial radius of $\rinit=5\pc$.
Figure~\ref{fig:tevol} plots time evolution of (a) radius $\rsnr$, (b)
deceleration parameter $\eta\equiv \vsnr t/\rsnr$, where $\vsnr$ is calculated
by finite differences of the measured $\rsnr$, and $\eta$ is smoothed, (c)
total $E_{\rm tot}$, thermal $E_{\rm th}$, and kinetic $E_{\rm kin}$ energies,
(d) mass of hot gas $\Mhot$ and shell gas $\Msh$, (e) total radial momentum
$\psnr$, and (f) pressure of hot gas $\Phot$.  The dotted and dashed lines
indicate theoretical predictions of each quantity for ST and idealized PDS
solutions as described in Section~\ref{sec:theory}, while the vertical
dot-dashed line denotes the predicted shell formation time from Equation
(\ref{eq:tsf}).  We also present in Figures~\ref{fig:rprof} and \ref{fig:snap}
the radial profiles and snapshots in the $\xhat$-$\yhat$ plane at $z=0$
(XY-plane) for selected times.  In Figure~\ref{fig:rprof}, blue dashed lines
show the ST self-similar solutions at $t=0.03\Myr$, while vertical dotted lines
are $\rsnr$ at each epoch.

At $t<\tsf$, the dynamical evolution of the SNR is characterized by the ST
solutions very well. The initial SNR in the simulation has a finite size and
zero kinetic energy, in contrast to an initial point source and a ratio $E_{\rm
kin}/E_{\rm th}=0.39$ of the ST solution.  However, as the shock expands into
the ambient medium, radial motions are rapidly generated.  After a transient of
only $t\sim 2\kyr$, all the physical quantities follow the ST solution.
Although the radial profiles shown in Figure~\ref{fig:rprof} differ from the ST
solutions at small radii $r<10\pc$ due to the finite initial size, they are
quite close near the shock where the most of mass and momentum are
concentrated.  As a consequence, there is excellent agreement with the ST
solutions for all integrated physical properties, as seen in
Figure~\ref{fig:tevol}. 

As the shocked hot gas cools, a thin and dense shell is formed (see green and
red lines of Figure~\ref{fig:rprof} and the 2nd and 3rd columns of
Figure~\ref{fig:snap}).  Shell formation is clearly distinguished by sudden
changes of several physical quantities. At shell formation, the expansion
stalls (sudden drop of $\vsnr$ and hence $\eta$ in Figure~\ref{fig:tevol}(b)),
the thermal energy and the mass of the hot gas begin to decrease
(Figure~\ref{fig:tevol}(c) and (d)), and the mass of the shell gas sharply
increases (Figure~\ref{fig:tevol}(d)).  We define the numerical shell formation
time, $\tsfnum$, as the time when $\Mhot$ attains its maximum\footnote{
\citet{1998ApJ...500...95T} use the time of maximum luminosity, $t_0$, to
identify shell formation; our definition based on the hot gas mass is in
practice quite similar.}.  Note that the theoretical prediction of $\tsf$ from
Equation (\ref{eq:tsf}) agrees with the maximum of $\Mhot$ fairly well (see
vertical line in Figure~\ref{fig:tevol}(d) and Table~\ref{tbl:uniform}). Since
the cooling function we use is different from the single power-law cooling
function used to derive Equation (\ref{eq:tsf}), $\tsfnum$ is slightly
different from $\tsf$, with slightly different dependence on $\ESN$ and $n_0$
as well.

After shell formation, the pressure of hot interior gas decreases much faster
than for the idealized adiabatic PDS solution, $\Phot\propto t^{-10/7}$ (dashed
line in Figure~\ref{fig:tevol}(f)).  This difference was reported by
\citet{1988ApJ...334..252C}, who found an additional $t^{-4/9}$ dependence in
thermal energy and hence the pressure of the hot gas (after shell formation).
Although the cooling time of the hot gas at small radii is still quite long
because of its low density, the density immediately inside the shell is higher
and this gas is able to cool.  In addition, the velocity interior to the shell
exceeds that of the shell itself (see red, cyan, and magenta lines in
Figure~\ref{fig:rprof}(d)), so that interior hot gas continues to accumulate,
condense, and cool at the inner surface (back) of the shell.  With a steadily
decreasing mass of hot gas, the interior pressure is ever lower than that of
the idealized PDS solution (which assumes a constant interior mass), adding
much less momentum (compare dashed and solid lines in
Figure~\ref{fig:tevol}(e)).  The evolution of radius with time nevertheless
seems to be roughly consistent with the PDS exponent $\eta=2/7$ (or $3/10$ for
the offset power-law fit in \citealt{1988ApJ...334..252C}; see
Figures~\ref{fig:tevol}(a) and (b)), but this can also be explained as a
transition from the ST to MCS phase, i.e. from $\eta=0.4$ to $\eta=0.25$
\citep{1998ApJ...500..342B,2004A&A...419..419B}.  Since the hot gas mass drops
very rapidly, the duration of the ``classical'' PDS stage is negligible.

As most of the hot gas has already piled into the shell by $t>$(2-3)$\tsf$, the
hot gas pressure is no longer high enough to drive the shell outward and add
appreciable momentum (cyan and magenta lines in Figure~\ref{fig:rprof}(b);
Figure~\ref{fig:tevol}(e)).  In addition, the shell becomes thicker at this
stage so that the thin shell approximation is no longer valid. The shell is
broken up in this high resolution simulation (fourth column of
Figure~\ref{fig:snap}) due to 
the pressure-driven thin shell overstability and/or
nonlinear thin-shell instability
\citep{1983ApJ...274..152V,1994ApJ...428..186V,1998ApJ...500..342B}.\footnote{
The instability of radiative shells was carefully studied by
\citet{1998ApJ...500..342B}.
Immediately after shell formation, a radiative reverse shock emerges,
resulting in a thin shell that is bounded by shocks on both sides.  This is
susceptible to the nonlinear thin shell instability
\citep{1994ApJ...428..186V}. For the low density case, similar to our fiducial
model SU-n1, however, \citet{1998ApJ...500..342B} showed that the reverse shock
soon becomes non-radiative and leaves the shell.  In this situation, the shell
is bounded by a forward shock and interior pressure, which can lead to the
pressure-driven thin shell overstability \citep{1983ApJ...274..152V}.}
Although we do not explicitly assign perturbations to our models, at the
initial interface between the SNR and ISM there are unavoidably grid-scale
nonlinear perturbations due to the mapping of a sphere to Cartesian
coordinates.  This seeds instability, which develops into strong corrugations.
Note that this instability is less prominent for realistic cases, in which the
multiphase structure of the ambient ISM dominates in shaping the shell (see
Section~\ref{sec:multi_TI}).

In models SU-n0.1, SU-n10, and SU-n100, we conduct analogous simulations to
model SU-n1, covering the range of ambient density of $n_0=0.1$ to $100$.  For
the high-resolution models with $n_0=0.1$, 1, 10, and 100, we set  $\Delta= 1$,
1/2, 1/4, and 1/8 pc, respectively, and $\rinit=4\Delta$; these choices are
based on our convergence study (see Section~\ref{sec:single_conv}).

Figure~\ref{fig:tevol_norm} plots the same quantities as in
Figure~\ref{fig:tevol}, but normalized using the corresponding numerical
measures at $\tsfnum$.  As before, this $\tsfnum$ is defined by the time when
the hot gas mass attains its maximum value. Note that $\tsfnum$, $\rsfnum$,
$\Msfnum$, and $\psfnum$ given in Table~\ref{tbl:uniform} for these
high-resolution models can be fitted by, respectively,
\begin{equation}\label{eq:tsfnum}
\tsfnum = 4.0\times10^4\yr\; n_0^{-0.59},
\end{equation}
\begin{equation}\label{eq:rsfnum}
\rsfnum = 22.1 \pc\; n_0^{-0.43},
\end{equation}
\begin{equation}\label{eq:Msfnum}
\Msfnum = 1550 \Msun\; n_0^{-0.29},
\end{equation}
and
\begin{equation} \label{eq:psfnum}
\psfnum = 2.00\times10^5 \momunit\; n_0^{-0.15}.
\end{equation}

Figure~\ref{fig:tevol_norm} demonstrates that the overall (normalized)
evolution is remarkably similar, irrespective of the ambient density. Since the
normalization factors are close to the analytic estimates from the ST solution
and the early evolution is expected to converge to the ST solution, this
agreement is trivial at $t<\tsfnum$.

However, even though the classical PDS solution is not realized in the
numerical models, the characteristic behavior after shell formation is
nevertheless essentially the same over a wide range of densities.  Model
evolution is thus ``congruent'' for varying ambient density when rescaled
relative to the SNR properties at $\tsfnum$.

In order to describe the transition between ST and PDS phases smoothly,
\citet{1988ApJ...334..252C} introduced an analytic description using an offset
power-law for radius and a complex fitting model for the thermal energy with
additional time dependence.  Since energy loss by radiative cooling begins
somewhat earlier than the shell formation, they also arbitrarily defined a PDS
time $t_{\rm PDS}\equiv\tsf/e$.  Here, instead of using the approach of
\citet{1988ApJ...334..252C}, we introduce a simpler analytic formula to
describe the post-shell formation phase separately.  As already seen in
Figure~\ref{fig:tevol}, the evolution of radius agrees reasonably well with the
classical PDS solution.  We thus simply adopt $\rsnr = \rsf(t/\tsf)^{2/7}$
(other exponents between 0.4 and 0.25 give similar results).  Following
\citet{1988ApJ...334..252C}, we describe the evolution of thermal energy after
shell formation with an additional time dependence as
\begin{equation}\label{eq:Eth}
E_{\rm th} = 0.8E_{\rm th,ST}\rbrackets{\frac{\rsf}{\rsnr}}^2\rbrackets{\frac{\tsf}{t}},
\end{equation}
where $E_{\rm th,ST} = 0.717 \ESN$ is the thermal energy at the ST stage, and
the factor ``0.8'' arises from precooling before the shell formation (see
dotted lines in Figure~\ref{fig:tevol_norm}(c)).  

The factor $\tsf/t$ is chosen to give a good match to the numerical solutions
for the time dependence of $E_{\rm th}$.  One might expect $E_{\rm th}$ to
decrease after shell formation proportional to the product of $\Mhot$ and the
interior temperature.  The mass-weighted mean temperature of hot gas is nearly
constant in this phase.  Thus, $\Mhot$ and $E_{\rm th}$ show similar time
dependence (approximately $\propto (t/\tsf)^{-1.64}$) until the thermal energy
of shell gas begins to dominate at $t>4\tsf$.

Using Equation (\ref{eq:Eth}), the effective pressure on the shell is then 
\begin{equation}\label{eq:Phot}
\Phot = \frac{E_{\rm th}}{2\pi \rsnr^3} = 0.8 \Psf
\rbrackets{\frac{\rsf}{\rsnr}}^5\rbrackets{\frac{\tsf}{t}}.
\end{equation}
We then have an expression for the momentum similar to Equation
(\ref{eq:ppds}):
\begin{equation}\label{eq:ppsf}
\psnr = \psf\sbrackets{1+0.53\int_1^{t_*} \frac{dt_*}{r_*^3t_*}} =
\psf\sbrackets{1+0.62(1-t_*^{-6/7})}.
\end{equation}
Note that with this modified, non-adiabatic PDS stage, the total momentum is
finite, $\psnr\rightarrow1.6\psf$ as $t\rightarrow\infty$; only 60\% of $\psf$
is added during the post shell formation stages.  The black dashed lines in
Figure~\ref{fig:tevol_norm} represent the modified PDS solution described
above.  The final momenta ($p_{\rm final}$) measured at $t=10\tsfnum$ are in
fact still smaller than $1.6\psf$ since no momentum is acquired after
$(4-5)\tsfnum$ as the hot gas pressure becomes comparable to the shell gas
pressure.  We find that $p_{\rm final}/\psfnum=1.49$, 1.46, 1.48, and 1.45 for
$n_0=0.1$, 1, 10, and 100, respectively.  In physical units, the final momentum
can be fitted as
\begin{equation}\label{eq:pfinal}
p_{\rm final}=2.95\times10^5\momunit n_0^{-0.16}.
\end{equation}

It is interesting to compare our numerical results with earlier studies based
on one-dimensional spherical models.  \citet{1998ApJ...500..342B} have measured
the shell formation time in terms of the ``transition'' time (Equation (3) of
their paper; simply defined by equating the cooling time with the age of the
SNR).  For $n_0=0.084$, 0.84, and $84$, their numerical measures of the shell
formation time are 1.35, 1.23, and 1.61 times our $\tsfnum$ for $n_0=0.1$, 1,
and 100, respectively.

A wide parameter space of density and metallicity has been covered by
\citet{1998ApJ...500...95T}. In their work, the numerical measures of the shell
formation time ($t_0$ in their notation), agree with our $\tsfnum$ within 20\%
for corresponding density. More interestingly, their ratios of total final
momentum (measured at $13t_0$) to the momentum at $t_0$ are 1.5-1.6 for
$n_H=0.1$ to $100\pcc$, similar to our findings.\footnote{In
\citet{1998ApJ...500...95T}, the total momentum is not explicitly provided.
Using the values presented in their Tables 1-4, we calculate the total momentum
as $\psnr\equiv(2M_R E_{R_{\rm kin}})^{1/2}$, where $M_R$ and $E_{R_{\rm kin}}$
are mass and kinetic energy of the SNR, respectively. Note that this definition
is consistent with direct integration of the specific momentum if most of the
mass is concentrated in a narrow region, which is true for both ST and
post-shell formation phases.} Even for more extreme parameters with
$n_H=10^{-3}$ and $10^3\pcc$ or different metallicities, the ratios still
remain around 1.7-2.

\subsection{Condition for Convergence}\label{sec:single_conv}

\begin{figure}
\epsscale{1.0}
\plotone{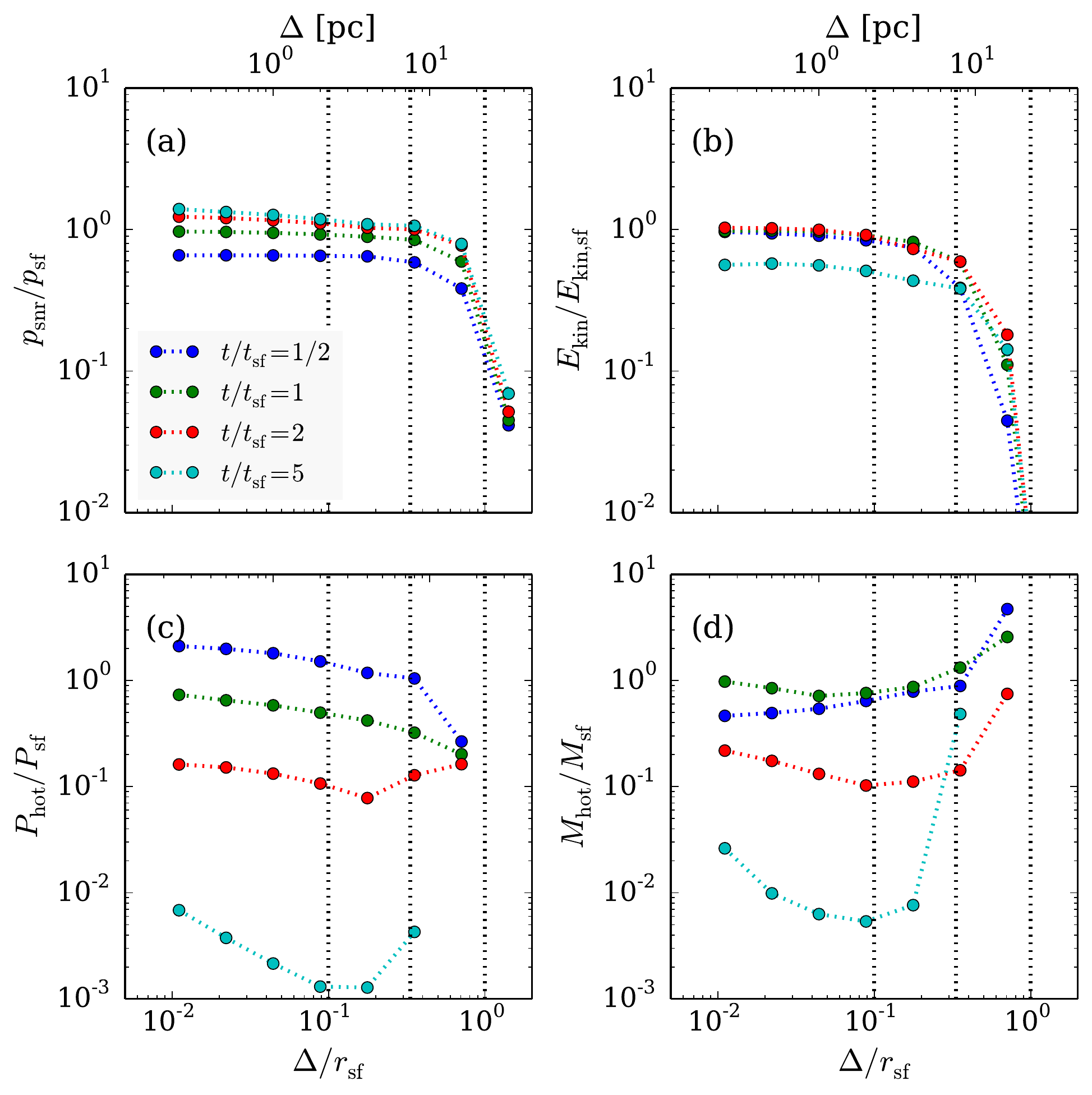}
\caption{
Convergence plot at varying grid resolution $\Delta$, showing (a) radial
momentum, (b) kinetic energy, (c) hot gas pressure, and (d) hot gas mass, for
model SU-n1.  Physical quantities are measured at $t/\tsf=1/2$ (blue), 1
(green), 2 (red), and 5 (cyan). All quantities are normalized based on the ST
solution at $\tsf$.  Top and bottom axes show grid scale in pc and relative to
$\rsf$.  For reference, vertical dotted lines denote $\Delta/\rsf=1/10,$ 1/3,
and 1.  Evidently, $\Delta/\rsf=1/10$ and $1/3$ are practical criteria for
robust and crude convergence, respectively.  \label{fig:dxnorm}}
\end{figure}

\begin{figure}
\epsscale{1.0}
\plotone{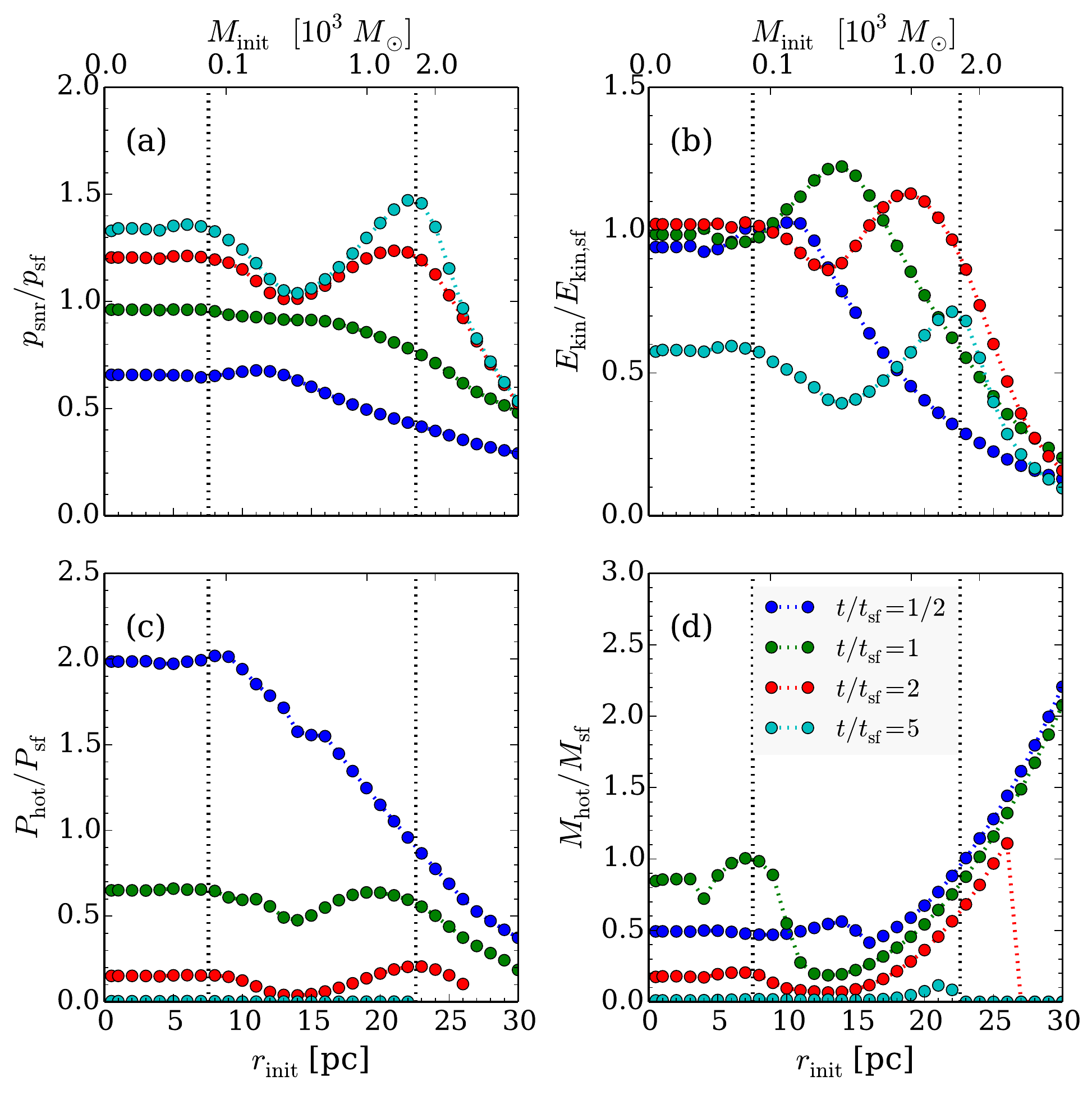}
\caption{
Convergence plot for varying initial SNR radius, $\rinit$, showing (a) radial
momentum, (b) kinetic energy, (c) hot gas pressure, and (d) hot gas mass, for
model SU-n1.  Physical quantities are measured at $t/\tsf=1/2$ (blue), 1
(green), 2 (red), and 5 (cyan).  All quantities are normalized based on the ST
solution at $\tsf$.  Top and bottom axes show $\Minit$ and $\rinit$ for the
SNR.  For reference, vertical dotted lines denote $\rinit/\rsf=1/3$ and 1.  All
tests use $\Delta =1/2\pc\sim 0.02\rsf$.     For all evolutionary stages and
all quantities, convergence is clear for $\rinit/\rsf<1/3$.  Although the final
momentum is within $\sim 30\%$ of the correct value for all cases with
$\rinit/\rsf<1$, the hot gas properties are far from correct  at $t<\tsf$ for
$1/3<\rinit/\rsf<1$.  \label{fig:rnorm}}
\end{figure}

\begin{figure}
\epsscale{1.0}
\plotone{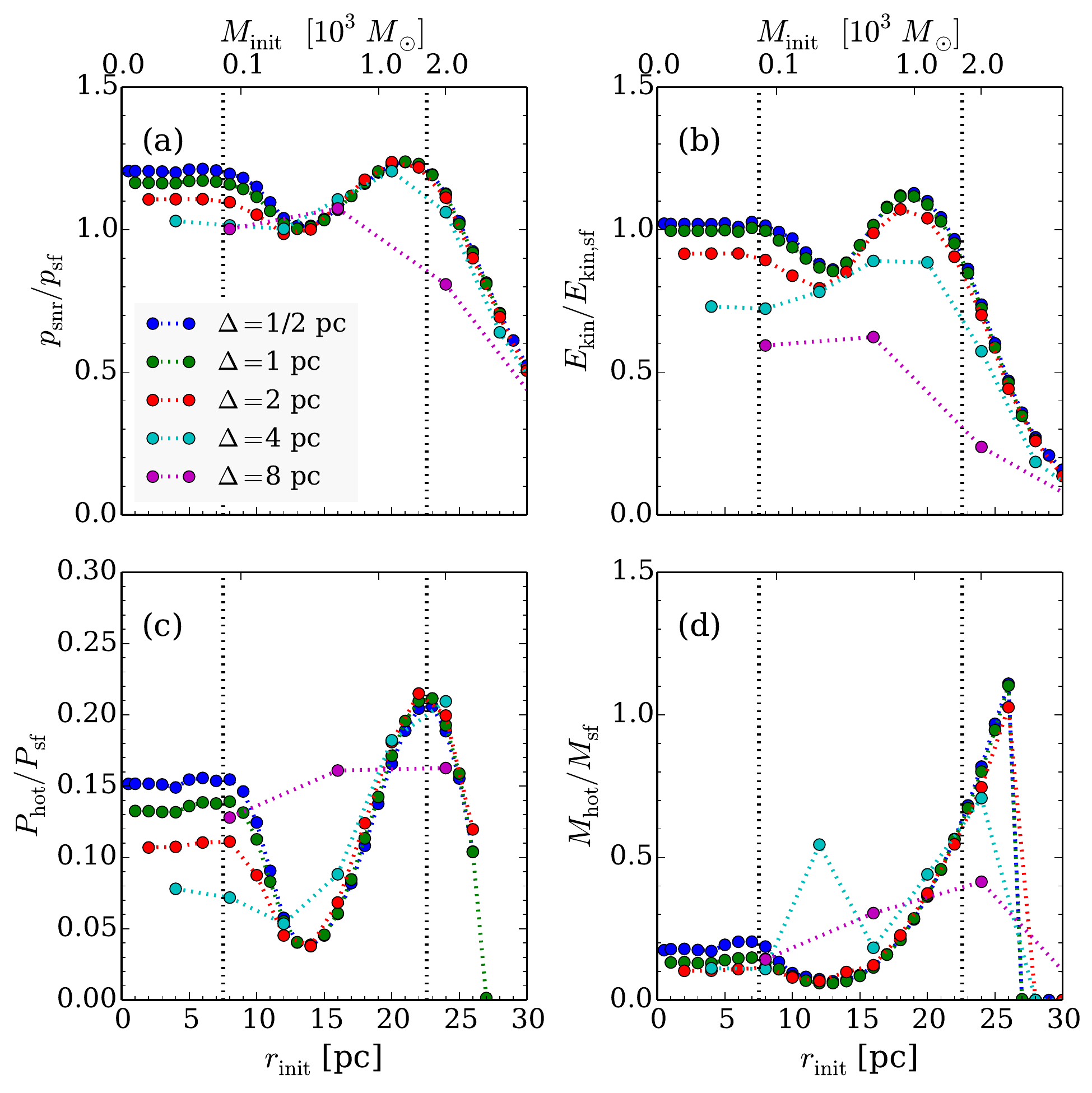}
\caption{
Same as Figure~\ref{fig:rnorm} but testing the effect of varying resolution
$\Delta$ at $t/\tsf=2$.  Convergence with respect to the initial SNR size is
satisfied for $\rinit/\rsf<1/3$ at all $\Delta$, but converged values fall
below the high-resolution benchmarks as the resolution gets poorer.
\label{fig:rnorm_dx}}
\end{figure}

\begin{figure}
\epsscale{1.0}
\plotone{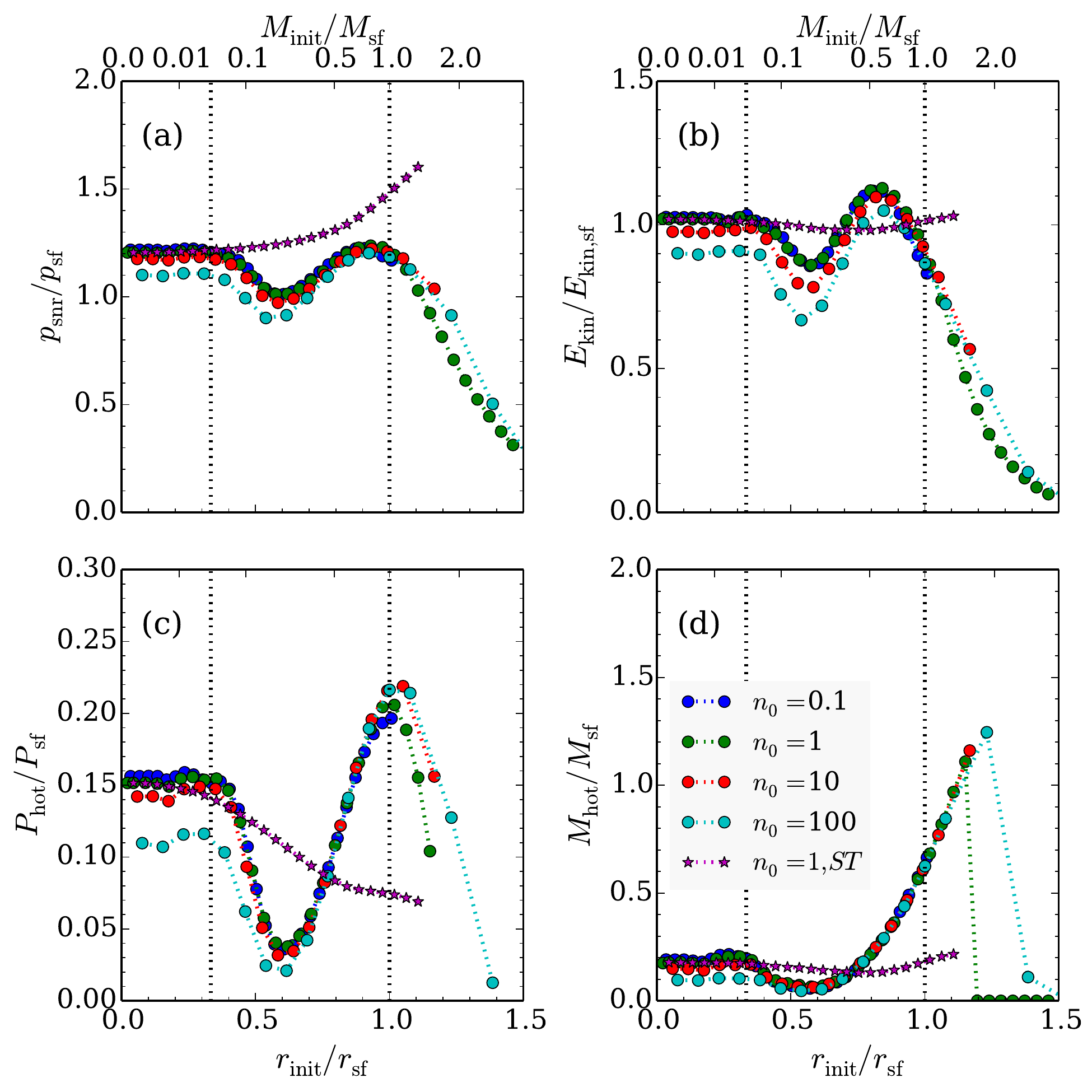}
\caption{
Same as Figure~\ref{fig:rnorm} but for different ambient density, and only
showing results for $t/\tsf=2$. The grid resolutions are $\Delta=1$, 1/2, 1/4,
and 1/8 pc for $n_0=0.1$, 1, 10, and 100, respectively, corresponding to
$\Delta/\rsf=0.015$, 0.02, 0.03, and 0.04. The bottom and top x-axes are
normalized by the radius and mass at the shell formation time for the
corresponding density.  In addition to models initiated with thermal energy
only, models initialized with Sedov-Taylor profiles are shown for $n_0=1$
(magenta stars).  Convergence requires $\rinit/\rsf<1/3$ irrespective of
ambient density.  \label{fig:rnorm_n}}
\end{figure}

As shown in the previous section, with sufficiently high resolution and a small
enough initial size for the SNR, we are able to reproduce in our 3D simulations
the evolution of radiative SNR as seen in previous 1D spherical models
\citep[e.g.,][]{1988ApJ...334..252C,1998ApJ...500...95T,1998ApJ...500..342B}.
Our models show a clear ST stage, but the post shell formation phase differs
from the ``classical'' PDS solution, similar to findings from previous
numerical work. Within $\sim 4\tsf$, the total radial momentum of the SNR
asymptotes to a value $\sim 1.5$ times the momentum at the time of shell
formation.  The above high-resolution SU simulations provide a benchmark, and
in this subsection, we repeat the SU simulations for a wide range of numerical
grid resolution ($\Delta$) and initial size of SNR ($\rinit$).  Our goal is to
determine minimum numerical requirements needed to reproduce the main features
of expanding SNRs on Cartesian grids, for the purposes of tracking SN feedback
effects in large-scale ISM and galactic models.  

To test dependence on grid resolution $\Delta$, we compare results for key
physical quantities in model SU-n1 at $t/\tsf=1/2$, 1, 2, and 5.
Figure~\ref{fig:dxnorm} plots (a) total momentum, (b) kinetic energy, (c) hot
gas pressure, and (d) hot gas mass, for a range $\Delta/\rsf = 0.01 $ to 1.5.
We assign $\rinit=\Delta$ as $\Delta$ varies, such that the initial thermal
energy of SN is distributed among eight zones nearest to the center of the
simulation domain (the SN explosion occurs at a grid corner).  All physical
quantities are normalized by the values of the ST solution at $\tsf$.  Here, we
use the analytic estimates of $\tsf$ and other variables rather than
numerically determined values since the numerical measures vary from model to
model, whereas for a resolution study we require a fixed reference.

For $\Delta/\rsf<1/10$ and 1/3, both the momentum and kinetic energy are
converged respectively within $20\%$ and $30\%$ of the highest-resolution
results at every evolutionary stage.  At late stages after shell formation, the
hot gas mass depends on the resolution since the mass is concentrated near the
boundary of the hot and shell gas (and is sensitive to the exact definition of
``hot'' vs ``shell'' at later times when the hot gas mass is small), but shows
very good convergence at $t/\tsf=1$ for $\Delta/\rsf<1/3$.  The convergence of
$\Phot$, which is volume weighted, is as good as $<30\%$ and $<50\%$ for
$\Delta/\rsf<1/10$ and 1/3, respectively, omitting $t/\tsf=5$ when the hot gas
has negligible effect on dynamics. For $\Delta>\rsf$, the SNR cools down
completely in a short time, so that there is no hot gas even at $t/\tsf<1/2$;
the momentum is $<10\%$ and kinetic energy is $< 1\%$ of the converged values.
Thus, the dynamical effect of SN feedback is impossible to model via injection
of thermal energy if the grid resolution exceeds $\rsf$ \citep[see
also][]{2014ApJ...788..121K}.  This is the origin of the ``overcooling''
problem cited in many numerical simulations of galaxy formation
\citep[e.g.,][]{1992ApJ...391..502K}, which at best resolve scales several tens
of pc and above.  

In addition to grid resolution, prescriptions for SN feedback in large-scale
simulations that are based on  thermal energy injection must also specify an
initial SNR radius, $\rinit$.  By varying $\rinit/\rsf$, we also have assessed
convergence with respect to this numerical parameter.  For these tests, we use
the SU-n1 model and adopt $\Delta=1/2\pc \sim 0.02\rsf$.
Figure~\ref{fig:rnorm} plots (a) radial momentum, (b) kinetic energy, (c) hot
gas pressure, and (d) hot gas mass, at several different times, as a function
of initial radius (or initial enclosed mass $\Minit$).  It is clear that
converged results are obtained at every evolutionary stages for $\rinit/\rsf <
1/3$.  Although momentum and kinetic energy that are not too far from correct
values can also be obtained with $\rinit\sim\rsf$, the evolutionary history and
internal profiles are completely different from the converged, resolved
solutions if $\rinit/\rsf>1/3$.

We have confirmed that the convergence criterion $\rinit/\rsf<1/3$ is valid for
different resolutions and ambient density.  Figures~\ref{fig:rnorm_dx} and
\ref{fig:rnorm_n} plot the same quantities as in Figure~\ref{fig:rnorm} at
$t/\tsf=2$, for different resolutions $\Delta$ and ambient density.  Similar
convergence trends are shown for lower resolution simulations and for a wide
range of varying ambient density.  With poorer resolution, all quantities
converged to smaller values than in the high-resolution benchmark models, as
seen in Figure~\ref{fig:dxnorm}.  If, instead of initializing with pure thermal
energy inside $\rinit$, the initial conditions instead employ the ST profile,
then the final momentum is converged out to  $\rinit/\rsf \sim 1/2$.  However,
initializing with the ST profile results in momentum exceeding the benchmark
value if $\rinit$ is comparable to or greater than $\rsf$.

We conclude that in order to obtain both convergence and accuracy (consistent
convergence), SNR must be initialized with both small enough radius
$\rinit/\rsf<1/3$, and high enough grid resolution.  The converged values
remain within roughly $25\%$ and $50\%$ of benchmarks for all physical
quantities with $\Delta/\rsf<1/10$ and 1/3, respectively.  In terms of injected
momentum, $\Delta/\rsf<1/10$ and 1/3 respectively give converged values within
5\% and 13\% of benchmarks at $t/\tsf=1$, and 18\% and 25\% of benchmarks at
$t/\tsf=10$.  The final momentum is marginally satisfactory when
$\rinit\sim\rsf$ and $\Delta/\rsf\sim1/3$ are adopted.  Any value $\rinit >
\rsf$, initialized with either pure thermal energy or a ST profile, gives poor
results. 

\section{Single SN in Two-Phase Cloudy Medium}\label{sec:single_2p}

\begin{deluxetable}{lccccccccc}
\tabletypesize{\footnotesize} \tablewidth{0pt} 
\tablecaption{Physical Quantities at the Shell Formation for S2P Models\label{tbl:2p}} 
\tablehead{ 
\colhead{Model} &
\colhead{} &
\colhead{$\bar n_H$} &
\colhead{$f_V$} &
\colhead{$f_M$} &
\colhead{$\tsf(\tsfnum)$} & 
\colhead{$\rsf(\rsfnum)$} & 
\colhead{$\Tsf$} &
\colhead{$\Msf(\Msfnum)$} & 
\colhead{$\psf(\psfnum)$} \\
\colhead{}&
\colhead{}&
\colhead{[$\pcc$]}& 
\colhead{}& 
\colhead{}& 
\colhead{[kyr]}&
\colhead{[pc]}&
\colhead{[$10^6\Kel$]}&
\colhead{[$10^3\Msun$]}&
\colhead{[$10^5\momunit$]}
}
\startdata 
\multirow{3}{*}{S2P-n0.1} 
& WHOLE & 0.1 & \nodata & \nodata  & 156(384)  & 59.5(83.3) & 0.31 & 3.08(3.44) & 2.94(3.21)\\
& CNM & 1.5      & 0.06 & 0.81 & 36    & 19  & 0.62  & 1.5  & 2.1 \\
& WNM & 0.017    & 0.92 & 0.15 & 420   & 130 & 0.19  & 4.9  & 3.7 \\
\hline
\multirow{3}{*}{S2P-n1}                                     
& WHOLE & 1   & \nodata & \nodata  & 43.7(100) & 22.6(31.5) & 0.57 & 1.68(1.44) & 2.17(2.13)\\
& CNM & 8.9      & 0.09 & 0.81 & 13    & 9.0 & 1.0   & 0.94 & 1.6 \\
& WNM & 0.14     & 0.84 & 0.12 & 130   & 52  & 0.34  & 2.8  & 2.8 \\
\hline
\multirow{3}{*}{S2P-n10}                                    
& WHOLE & 10  & \nodata & \nodata  & 12.2(24.3) & 8.56(10.7) & 1.04 & 0.92(0.70) & 1.60(1.46)\\
& CNM & 110      & 0.08 & 0.83 & 3.3   & 3.2 & 1.9   & 0.49 & 1.2 \\
& WNM & 1.5      & 0.88 & 0.13 & 35    & 19  & 0.64  & 1.5  & 2.1 \\
\hline
\multirow{3}{*}{S2P-n100}                                   
& WHOLE & 100 & \nodata & \nodata & 3.43(5.30) & 3.25(3.63) & 1.90 & 0.50(0.36) & 1.18(1.00)\\
& CNM & 1300     & 0.07 & 0.82 & 0.85  & 1.1 & 3.7   & 0.26 & 0.85\\
& WNM & 17       & 0.91 & 0.15 & 9.2   & 6.9 & 1.2   & 0.80 & 1.5 
\enddata
\tablecomments{ 
Column 1: model name.
Columns 3-5: mean hydrogen number density (whole medium - top row, 
CNM - middle row, WNM - lower row); volume and mass fractions 
measured at saturation of TI.
Columns 6-10: shell formation predicted 
time, radius, postshock temperature, 
swept-up mass, and total momentum for propagation into a uniform medium
with density $\bar n_H$. 
See Equations (\ref{eq:tsf}), (\ref{eq:rsf}), (\ref{eq:Tsf}),
(\ref{eq:Msf}), and (\ref{eq:psf}), respectively.
Numerical measures at $\tsfnum$ averaged over 10 realizations of each model are
shown in parentheses in the ``WHOLE'' rows.
}
\end{deluxetable}

\begin{figure}
\epsscale{1.0}
\plotone{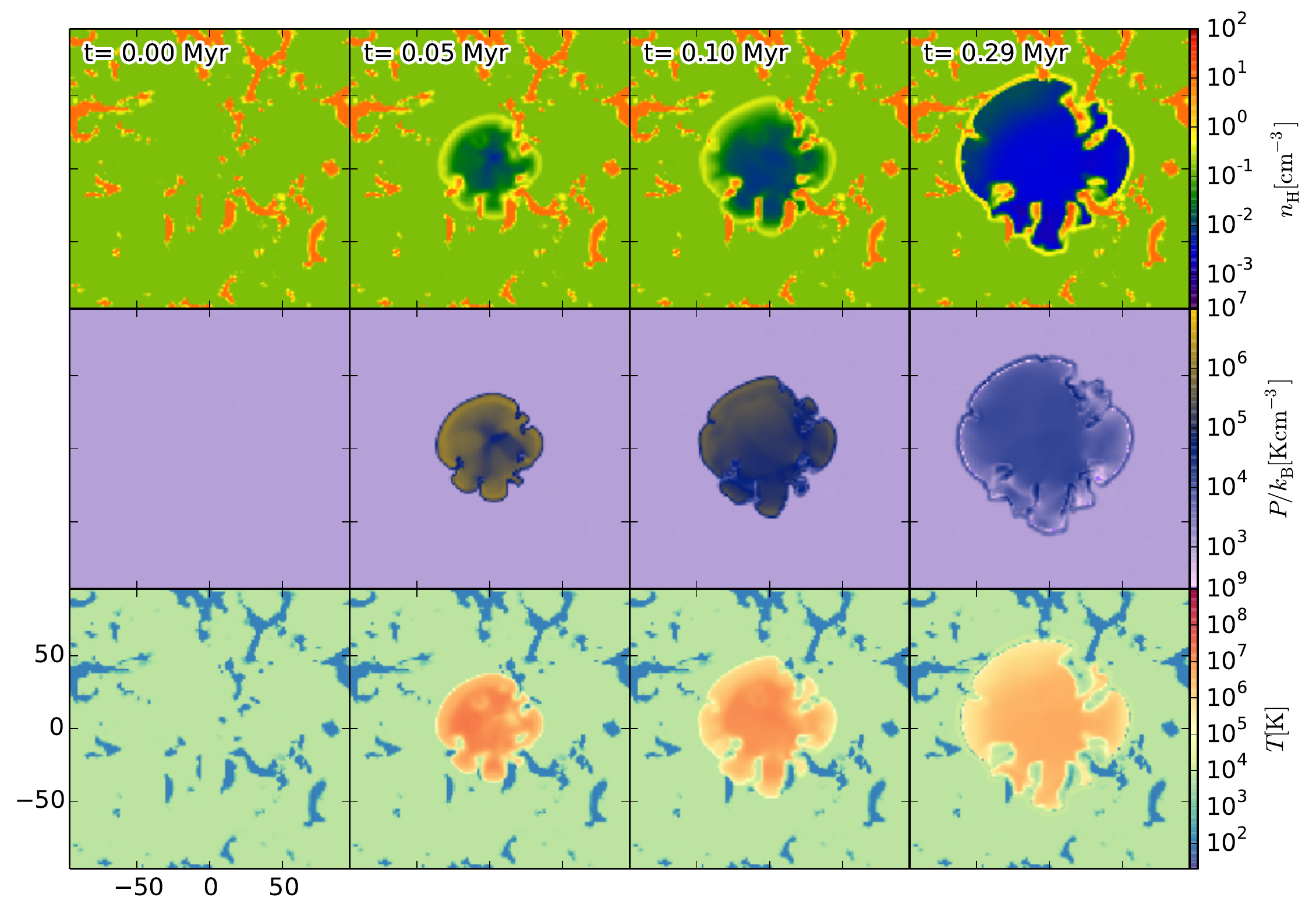}
\caption{
Example XY-slices of model S2P-n1 (two-phase initial medium with $n_0 = 1$) for
grid resolution $\Delta = 1.5\pc$.  From top to bottom, logarithmic color
scales show number density, pressure, and temperature. 
\label{fig:snap_2p}}
\end{figure}

\begin{figure}
\epsscale{1.0}
\plotone{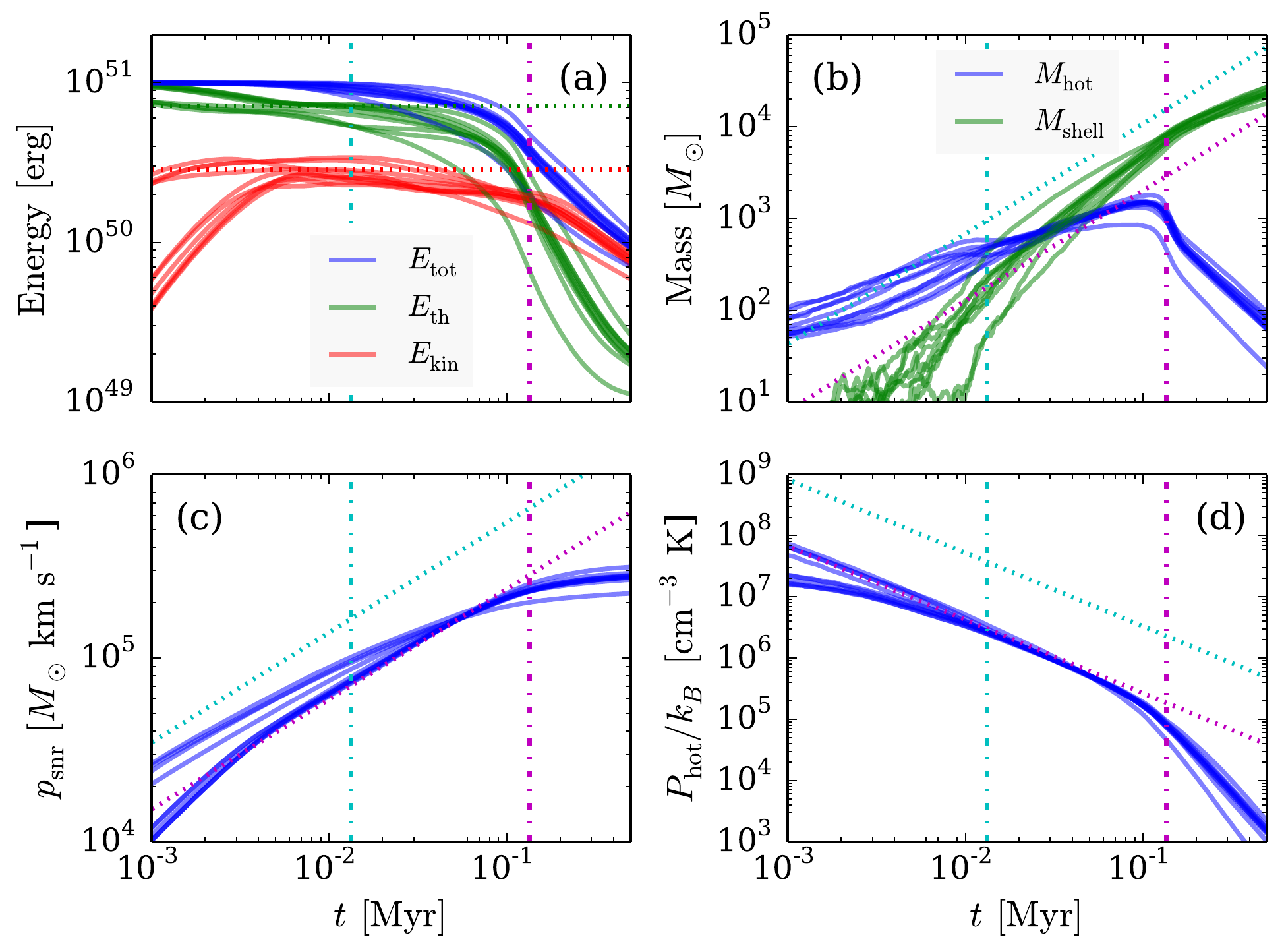}
\caption{
Time evolution for 10 different realizations of model S2P-n1. Shown are (a)
total, thermal, and kinetic energies, (b) hot and shell gas masses, (c) total
radial momentum, and (d) hot gas pressure.  Vertical dot-dashed lines are the
corresponding expected shell formation time for a uniform medium at the  CNM
(cyan) and WNM (magenta) density.  Dotted lines are the ST solution for $n_{\rm
CNM}$ (cyan) and $n_{\rm WNM}$ (magenta).  \label{fig:tevol_2p}}
\end{figure}

\begin{figure}
\epsscale{1.0}
\plotone{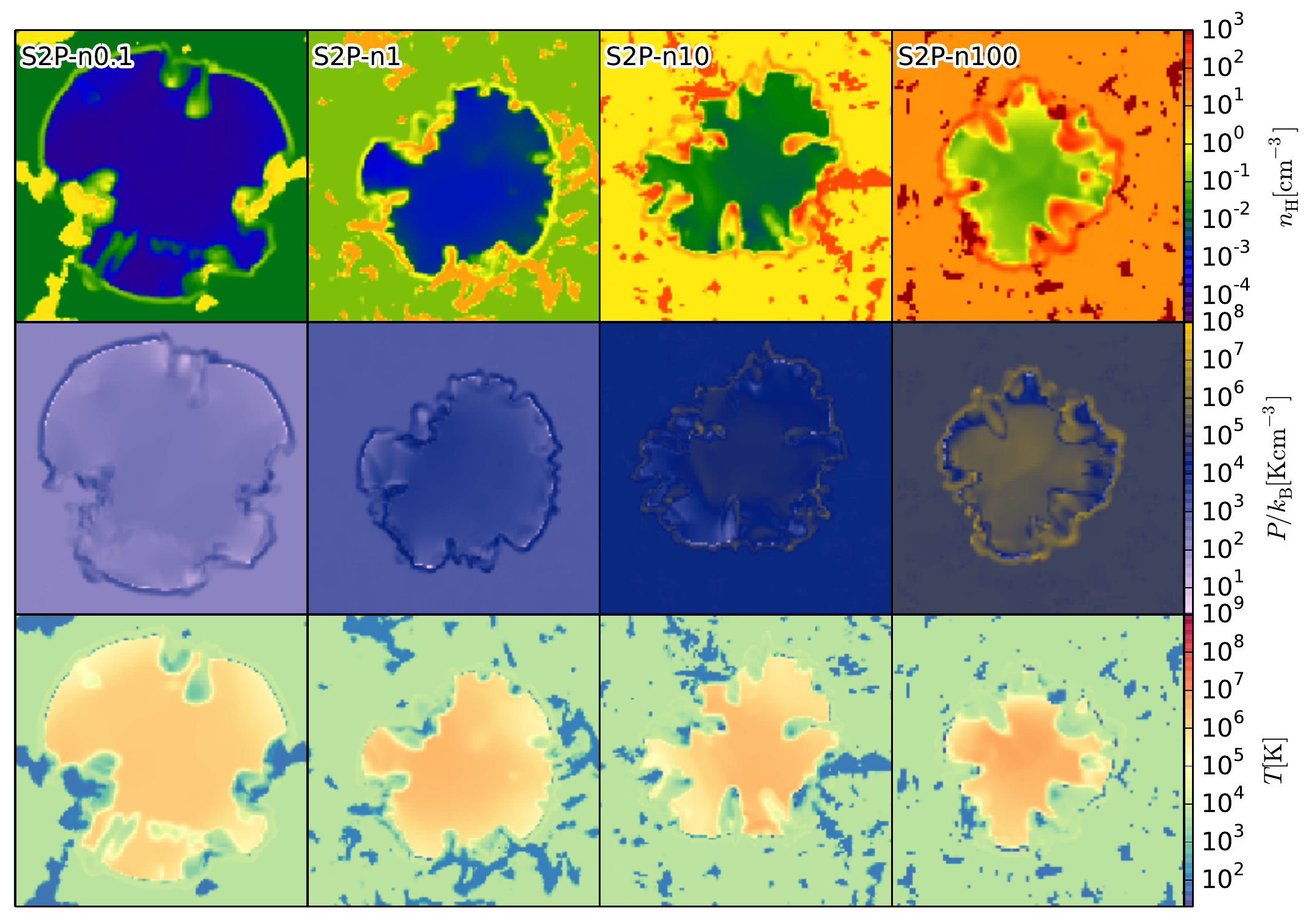}
\caption{
Example XY-slices of all S2P models at $t/\tsf\sim10$.  The dimensions of the
slices shown  $384\pc\times384\pc$ for S2P-n0.1, $192\pc\times192\pc$ for
S2P-n1, $64 \pc\times64 \pc$ for S2P-n10, and $24 \pc\times24 \pc$ for
S2P-n100.  The grid resolutions are $\Delta = 3$, 1.5, 0.5, and 0.25 pc for
$n_0=0.1$, 1, 10, and 100.  From top to bottom, logarithmic color scales show
number density, pressure, and temperature. 
\label{fig:snap_s2p}}
\end{figure}

\begin{figure}
\epsscale{1.0}
\plotone{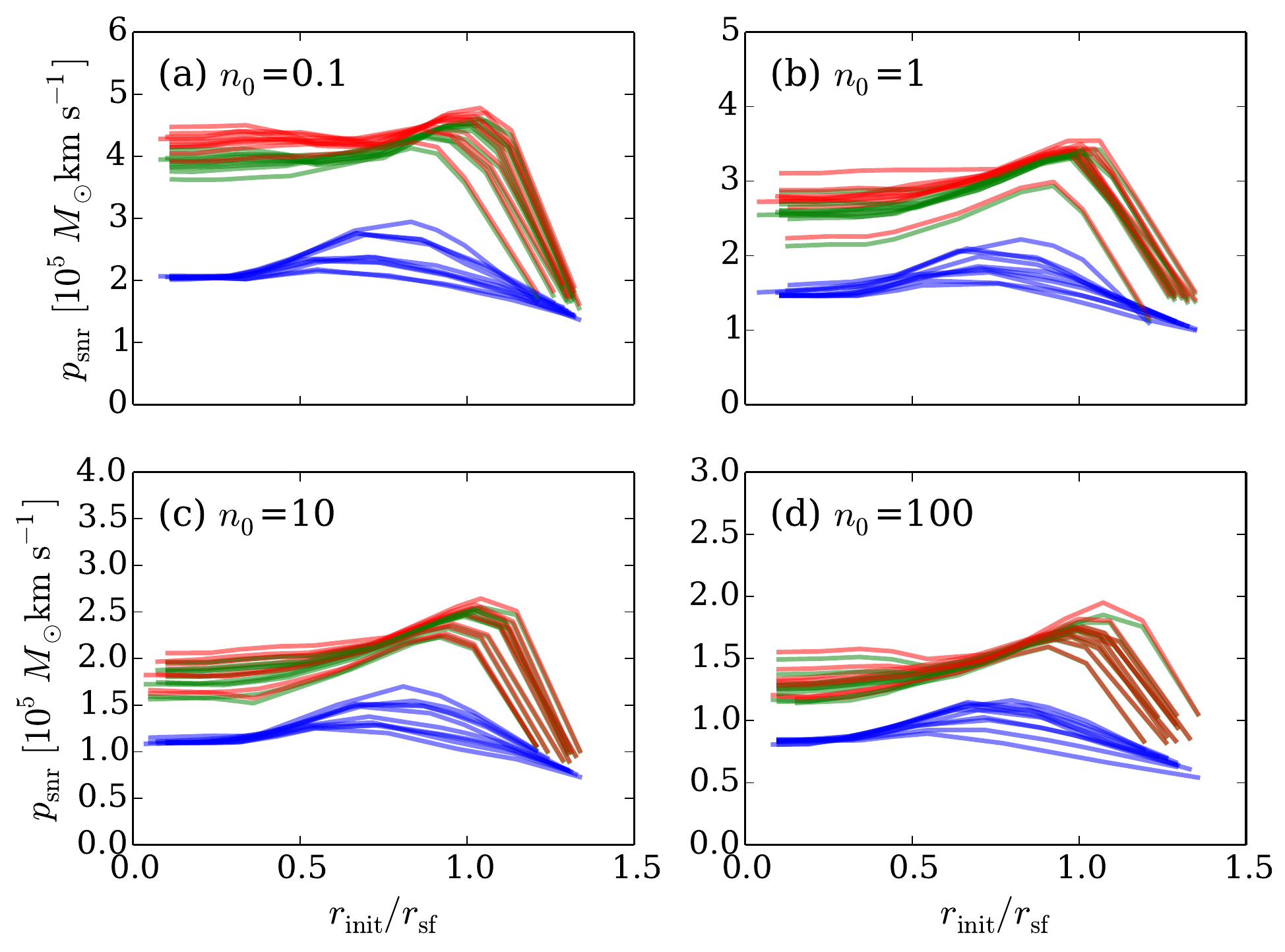}
\caption{
Convergence plots of total momentum as a function of the normalized initial
radius of the SNR for S2P at $t/\tsf=1$ (blue), 5 (green), 10 (red).  The same
condition derived from uniform background ($\rinit/\rsf<1/3$;
Section~\ref{sec:single_conv}) can be applied here for ``consistent
convergence''.
\label{fig:rnorm_2p}}
\end{figure}

In the real ISM, the ambient gas around SNe is not uniform, but clumpy.  The
immediate environment will be a photoionized \ion{H}{2} region that is also
impacted by the progenitor's stellar winds.  At larger scales, comparable to
$\rsf$, SNe may propagate either into dense molecular gas of a parent cloud (if
it has not yet been dispersed), into the cloudy warm/cold atomic medium, or
into very low density hot gas produced by previous generations of SNRs.  The
case of propagation into the warm/cold atomic ISM is especially important to
quantify, as SNRs are believed to be the main source of turbulence in this
phase of the ISM.

The atomic ISM ($T<10^4\Kel$) is regulated by two dominant cooling processes,
atomic fine structure line cooling and Ly$\alpha$, yielding a two-phase medium
with temperature near $10^2\Kel$ and $10^4$, respectively; grain-photoelectric
net heating dominates nearly irrespective of the gas temperature
\citep{1995ApJ...453..673W,2003ApJ...587..278W}.  For a range of pressures,
there is a third possible equilibrium temperature in between the warm and cold
phases, but this has relatively low occupation due to the thermal instability,
with a growth time $\sim {\rm Myr}$ that is shorter than large-scale dynamical
times in the ISM
\citep{1965ApJ...142..531F,1969ApJ...155L.149F,1995ApJ...453..673W,2004ApJ...601..905P,2008ApJ...681.1148K}.
This results in an ISM structure consisting of cold, dense clouds embedded in a
background of warm, rarefied gas.  The warm and cold medium densities differ by
two orders of magnitude.

To study propagation of SNRs into atomic ISM gas, we first construct two-phase
``background states'' via nonlinear simulations of the thermal instability (TI
runs). The initial conditions for each TI run is a thermally unstable
equilibrium with $10\%$ density perturbations.  For each different mean density
($n_0=0.1-100$), $\Gamma \propto n_0$ (as in the SU models; this scaling is
consistent with self-regulated star formation), which makes the initial state
thermally unstable.\footnote{Conditions at the highest density $n_0=100
\cm^{-3}$ are more likely to be found in the molecular ISM than the atomic ISM.
However, as the chemical state of the ambient gas is unimportant when shocks
are strong, it is useful to consider this case as representative of the
high-density, clumpy conditions that are present in the centers of normal disk
galaxies and more pervasively in starbursts.} We follow each TI run long enough
to obtain a saturated two-phase state with pressure balance.  We then explode a
SN at the center of the simulation domain. 

As the details of the cloudy structure that a SNR encounters affects its
propagation, for each mean density in the S2P model set we use 10 different
realizations of the cloudy background state (from TI runs with different
perturbation seeds).  Table~\ref{tbl:2p} lists from 3rd to 5th columns the mean
densities ($\bar n_H$), volume fractions ($f_V$), and mass fractions ($f_M$) of
CNM ($T<184\Kel$) and WNM ($T>5050\Kel$) gas in the saturated state, averaged
over these 10 realizations.  The temperature cuts are determined by the maximum
and minimum temperatures of the stable CNM and WNM, respectively, for our
choice of cooling function
\citep[see][]{2002ApJ...564L..97K,2008ApJ...681.1148K}.  Variations of the mean
values in different realizations are less than 1\%.  Note that the sum of $f_V$
and $f_M$ is not 1 since a small fraction of gas resides in the cloud
interface.  As $\Gamma \propto n_0$, the volume and mass fractions of the two
phases are similar at all $n_0$, and the densities in each phase are
proportional to $n_0$.  Table~\ref{tbl:2p} also lists for reference the
physical quantities at shell formation that would apply for the ST solution
propagating into {\it uniform} high-density (cold cloud) gas, {\it uniform} low
density (intercloud) gas, and gas at the average density of the simulation.
Numerical measures of the quantities averaged over 10 realizations are also
shown in parentheses.  As before, we define $\tsfnum$ as the time when the hot
gas mass reaches a maximum.

Power-law fits for averaged numerical measures are given by
\begin{equation}\label{eq:tsfnum_2p}
\tsfnum = 9.6\times10^4\yr\; n_0^{-0.62},
\end{equation}
\begin{equation}\label{eq:rsfnum_2p}
\rsfnum = 30.2 \pc\; n_0^{-0.46},
\end{equation}
\begin{equation}\label{eq:Msfnum_2p}
M_{\rm sf,hot}^{\rm n}= 1540 \Msun\; n_0^{-0.33},
\end{equation}
and
\begin{equation} \label{eq:psfnum_2p}
\psfnum = 2.16\times10^5 \momunit\; n_0^{-0.17}.
\end{equation}
Note that the above is the mass of {\it hot} gas at $\tsfnum$; there is several
times as much ``shell'' gas because initially-cold, dense clouds are
accelerated by the shock and by the expanding gas in the interior of the SNR.
Compared to the uniform-medium case (Equations
(\ref{eq:tsfnum})-(\ref{eq:psfnum})), dependencies on the mean ambient density
are similar.  Although the shell formation epoch $\tsfnum$ is delayed for a
two-phase medium, the hot gas mass and total momentum at shell formation agree
very well with the numerical measures of these quantities for a SNR in a
uniform background (SU models), as well as analytic estimates.

Figure~\ref{fig:snap_2p} displays snapshots in the XY plane of an example run
for model S2P-n1. The time marked in the top row is the time since the
explosion. For this run, we set the grid resolution to $\Delta=1.5\pc$ and the
initial size of SN to be $\rinit/\rsf\sim 0.2$, for an  $\rsf$ estimated using
the mean density. For this specific example, the saturated two-phase ISM is
characterized by $(n_H, T) = (8.8\pcc, 106\Kel)$ and $(0.14\pcc, 6760\Kel)$ for
CNM and WNM, respectively.  The simulation volume is mainly occupied by the WNM
with $f_V=0.84$, while the majority fraction $f_M=0.81$ of the mass resides in
the CNM.  Since the SNR expands solely through the WNM in most directions, the
overall size of SNR is similar to expectations for a uniform medium with
$n_0=0.14$ (especially in the 2nd quadrant of the XY plane in
Figure~\ref{fig:snap_2p}).  In some directions, the blastwave encounters small
CNM clouds; these may be 
destroyed by the passage of the shock and/or form pillars within the
SNR.\footnote{ 
The ablation of clouds by dynamical instabilities is limited because of
insufficient resolution. In addition, dense clouds may also evaporate after the
shock passes if conduction from the surrounding hot medium is large enough; the
present simulations do not include explicit conduction. Both effects are not
fully captured, but see discussion in Section~\ref{sec:sumndis}.}
Large CNM clouds are also shock compressed, and may partly block the SNR
expansion, but over time they are also carried outward.  At late times, the
shell morphology is mainly shaped not by the shell instability observed in
Figure~\ref{fig:snap} but by the initial structure of the ambient two-phase
medium.

For S2P-n1 (10 different realizations), Figure~\ref{fig:tevol_2p} plots time
evolution of (a) total, thermal, and kinetic energies, (b) hot and shell gas
masses, (c) radial momentum, and (d) hot gas pressure.  Also indicated with
vertical dot-dashed lines are the ST solutions for the shell formation time
when propagating into the CNM density (cyan; $\tsfc$) and WNM density (magenta;
$\tsfw$).  Dotted lines in panels (b), (c), (d) give ST solutions for an
ambient medium that is 100\% CNM (cyan) and 100\% WNM (magenta).  Beginning at
$\sim \tsfc$, part of the SNR's energy is radiated away as the blastwave
envelopes and shocks CNM clouds; both thermal and kinetic energy are reduced
below the ST values.  Since the most of volume is filled with the WNM, however,
all physical quantities except the mass remain close to those of the ST
solution in a uniform WNM until $\sim \tsfw$. If there were no PDS stage, the
momentum injected to a two-phase medium would be expected to lie  between
$\psf$ for $\bar n_{\rm CNM}$ and $\bar n_{\rm WNM}$ (last column of
Table~\ref{tbl:2p}).  As $\psf$ is insensitive to the ambient medium density
($\psf\propto n_0^{-0.13}$ predicted from Equation (\ref{eq:tsf}), or
$\psf\propto n_0^{-0.17}$ using the numerical results in
Table~\ref{tbl:uniform}), and the momentum only modestly increases during the
PDS stage, the total injected momentum by a single SN will always be of order
$\psf$ for the mean density of the medium, $n_0$. 

Figure~\ref{fig:snap_s2p} displays example XY-slices for S2P models with
different mean density at $t/\tsf\sim10$.  Although the initial cold and warm
volume fractions are similar for all models, the cloud structures are
different; there are more small clouds in higher density models. For the
present models, the difference in structure at TI saturation is caused by
differences in TI growth rates; smaller clouds can grow faster with higher
heating rates in higher density models.\footnote{ Additional physical
ingredients ignored here such as turbulence, thermal conduction, and magnetic
fields may also play an important role in shaping the ISM structure and in
particular the size spectrum of cold clouds.} The effective contribution from
dense clouds increases with background density because of the increasing total
cloud cross-section when there are more clouds. Thus, the numerical measures of
physical quantities (in parentheses of Table~\ref{tbl:2p}) are closer to the
analytic estimates for CNM as the mean density gets higher.  However, the total
injected momentum remains similar to the estimates from the mean density.

For ten different realizations at each mean density $n_0$,
Figure~\ref{fig:rnorm_2p} plots the radial momentum at $t/\tsf=1$ (blue), 5
(green), and 10 (red), where $\tsf$ is computed using $n_0$.  Similar to Figure
\ref{fig:rnorm}, we show the result of varying the initial radius of the SNR.
At each $n_0$, we use a reference value of $\rsf$ based on Equation
(\ref{eq:rsf}) using $n_0$ (note, however, that the actual mean density within
this $\rsf$ will differ from $n_0$ due to the particular realization of the
cloudy structure).  The resolution $\Delta$ is set to be smaller than
$\rsf/10$, which we have found (using tests similar to those of section
\ref{sec:single_conv}) is sufficient for convergence.  The horizontal axis is
normalized by radius $\rsf$ using $n_0$ (see Equation (\ref{eq:rsf})).  The
final momentum converges provided $\rinit<\rsf/3$, similarly to what we found
in the uniform medium case.  This ``consistent convergence'' criterion
guarantees convergence at all evolutionary stages; convergence at late stages
only is possible up to slightly larger $\rinit$.  For $\rinit > \rsf$, the
final momentum is significantly lower than the converged value.    

For mean number density of $n_0=0.1$, 1, 10, and 100, the mean values of the
momentum at $t/\tsfnum=10$ are $p_{\rm final}/\psfnum=1.33$, 1.30, 1.29, and
1.32, respectively.  Similarly to Equation (\ref{eq:pfinal}), the final
momentum in the physical units can be fitted by
\begin{equation}\label{eq:pfinal_2p}
p_{\rm final}=2.8\times10^5\momunit n_0^{-0.17}.
\end{equation}
The final momentum is always of order of $\psf$, but slightly smaller for a
clumpy medium than in the uniform-medium case.  For different realizations of
the model at each mean density (i.e. different seeds for the random
perturbations), there are only 20-40\% variations in the final momentum. 

\section{Multiple SNe in Two-Phase Medium}\label{sec:multi_TI}

\begin{figure}
\epsscale{1.0}
\plotone{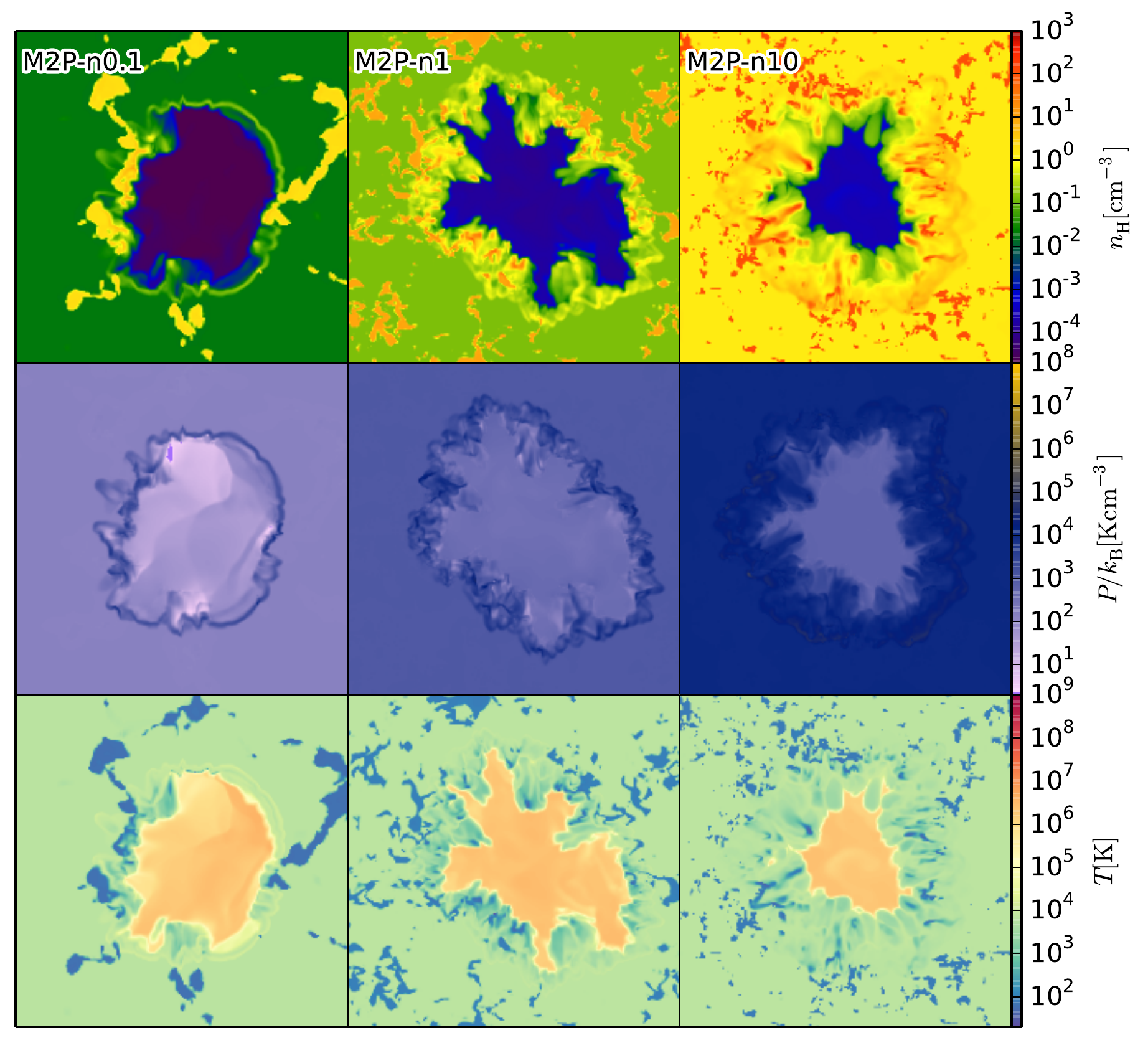}
\caption{
Example XY-slices of all M2P models at $t=2\Myr$.  The dimensions of the slices
shown are $768\pc\times768\pc$ for M2P-n0.1, $384\pc\times384\pc$ for M2P-n1,
and $192\pc\times192\pc$ for M2P-n10, The grid resolutions are $\Delta = 4$,
1.5, and 0.75 pc for $n_0=0.1$, 1, and 10.  From top to bottom, logarithmic
color scales show number density, pressure, and temperature.
\label{fig:snap_m2p}}
\end{figure}

\begin{figure}
\epsscale{1.0}
\plotone{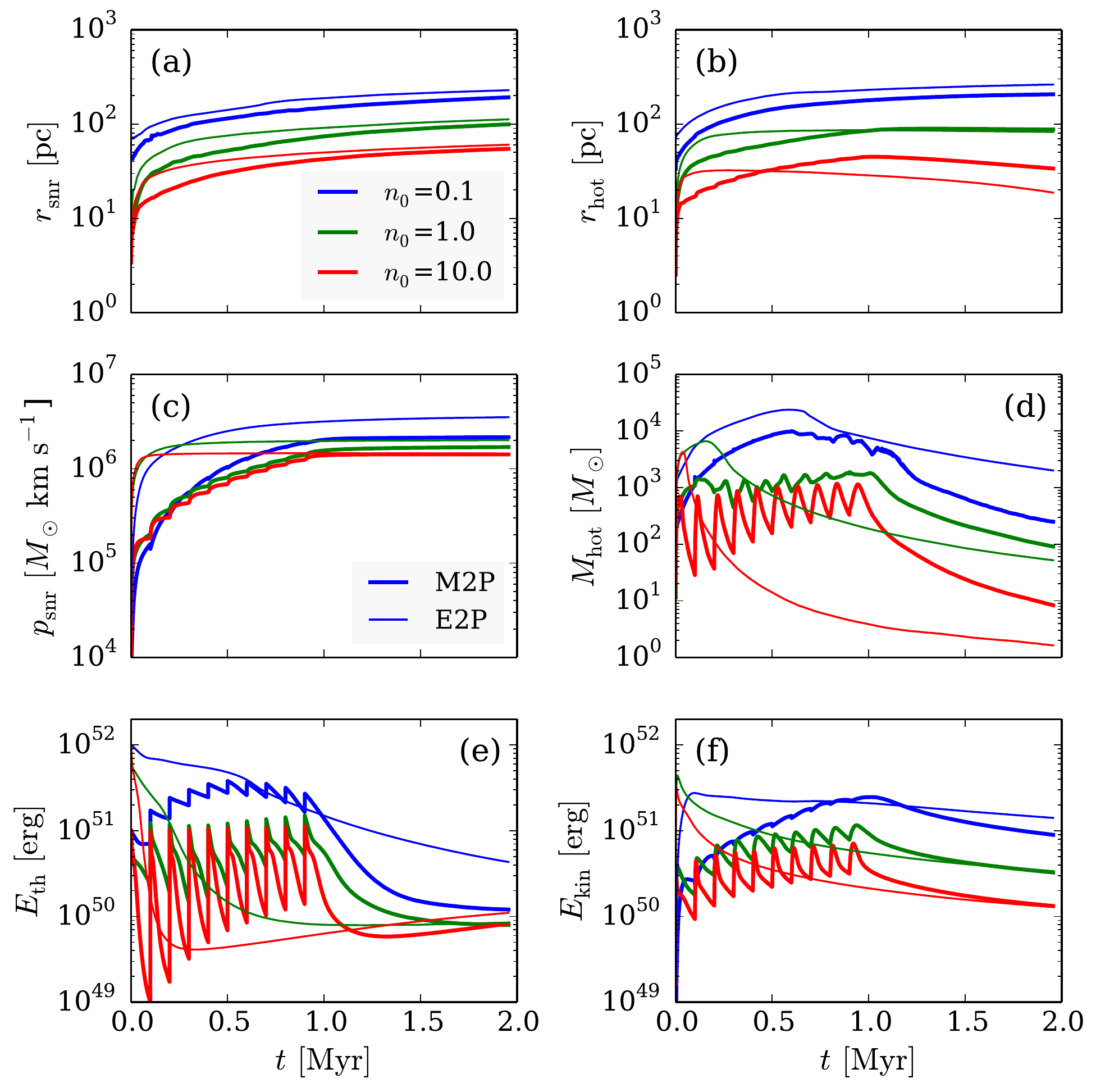}
\caption{
Time evolution of models with multiple SNe (M2P models, shown in thick lines),
with SN interval interval $\tsn=0.1\Myr$ and $\ESN=10^{51}{\;\rm erg}$. Panels
show (a) the mass-weighted mean radius of the combined SNR $\rsnr$, (b) the
effective  radius of the hot gas volume $r_{\rm hot}\equiv (3V_{\rm
hot}/4\pi)^{1/3}$, (c) radial momentum, (d) hot gas mass, (e) thermal energy,
and (f) kinetic energy.  For comparison, single-event models (E2P, with
$\ESN=10^{52}{\;\rm erg}$) are shown with thin lines.  The mean density of the
background ambient gas is $n_0=0.1$ (blue), 1 (green), and 10 (red).  Multiple
SNe become relatively less efficient at injecting momentum when the shell
formation time is shorter, at higher ambient density.
\label{fig:tevol_multi}}
\end{figure}

In Sections \ref{sec:single_ref}, \ref{sec:single_conv}, and
\ref{sec:single_2p}, we considered the expansion of a single SNR in either a
uniform or a clumpy medium for a wide range of mean densities ($n_0=0.1 -100
\cm^{-3}$), covering conditions that would apply in the atomic or molecular
ISM. There, we evaluated the momentum, mass of hot gas, and other basic
properties of the SNR, and determined numerical requirements for consistent
convergence, i.e. criteria such that simulation results agree with
high-resolution benchmarks, at times before and after the SNR becomes strongly
radiative.  In this section, we extend to the case of multiple SNe that
sequentially explode and expand into a surrounding two-phase medium.  For
multiple SNe, many more defining parameters enter the problem setup than for
the case of a single SN, so a comprehensive study will be deferred to future
work.  However, it is useful to explore a few simple cases to assess potential
differences  in the momentum injection per event, compared to that from a
single SN.
Of course, the momentum injection from multiple spatially correlated SNe
is also subject to global effects, since the superbubble that forms can expand
beyond the scale height of the surrounding ISM and break out of the disk,
venting into the galactic halo
\citep[e.g.,][]{1989ApJ...337..141M,1990A&A...237..207T,2012MNRAS.421.3522H}
and limiting further expansion within the ISM.  Here, we do not consider these
global effects (but see Section~\ref{sec:sumndis} for further discussion).

We consider the case in which SNe occur at intervals of 0.1Myr; this could
represent the effects of a moderate-mass cluster ($M_{\rm cluster}\sim 3 \times
10^4 \Msun$) in which the event rate is not sufficient for energy deposition to
reach the continuous limit.  We consider a range of mean densities $n_0=1-10$
in the ambient medium, and as for the S2P models we create a ``background''
two-phase state by allowing the TI to reach saturation.  

For the single-SN models, the initial size of the SNR was set relative to
$\rsf$, the shell formation radius at the mean density of the ambient medium.
Although for our tests $n_0$ has a pre-determined value and it is trivial to
set $\rinit$ directly, more generally the ambient density varies and it is more
practical to define the initial SNR region based on its total enclosed mass.
We take this approach for the M2P models.  To obtain an equivalent condition to
$\rinit/\rsf<1/3$, we begin with $\rinit=\Delta$ (thermal energy is distributed
in eight zones closest to the center) and calculate the enclosed mass
$\Minit\equiv \sum_{r<\rinit} \rho \Delta^3$ and the mean density
$\rhoamb=\Minit/V_{\rm init}$, where $V_{\rm init} = \sum_{r<\rinit}\Delta^3$.
If $\Minit/\Msf<1/27$ (where $\Msf$ is obtained with Equation (\ref{eq:Msf}) for
this computed $\rhoamb$) we increase $\rinit$ by $\Delta/2$ and repeat the
calculation in order to get the largest possible initial size of SNR that is
compatible with the convergence condition $\rinit/\rsf <1/3$ (i.e.
$\Minit/\Msf<1/27$).

For each SN event, once a sphere with appropriate $\rinit$ is defined, we
redistribute the enclosed  mass, momentum, and internal energy uniformly within
$\rinit$, and assign additional thermal energy adding up to $\ESN$. This
prescription prevents setting the temperature to extremely high values for very
rarefied regions produced by previous SN explosions.\footnote{A more realistic
prescription, starting from a free expansion phase with ejecta, would be
possible for subsequent SNe expanding into hot regions, because the first SN
makes the ambient density low enough to resolve this phase on a grid that is
suitable for large-scale ISM models.  We defer explorations of this and other
alternatives to future work, keeping in mind that complex ``early'' feedback
such as stellar winds and  \ion{H}{2} region expansion may also alter the
ambient state for secondary SNe.  Here, we simply adopt a thermal energy
initialization that is as similar as possible to our treatment of single SNR,
in order to compare the momentum that is injected.} All SNe explode at the
center of the simulation domain.

For our M2P model set, we run each simulation until 10 SNe, with
$\ESN=10^{51}\erg$ each, have exploded.  For reference, we also run comparison
simulations with a single explosion of energy $\ESN=10^{52}\erg$ (E2P).  After
the last SN (at $t=0.9\Myr$), we run each simulation out until $t=2\Myr$ to
measure the final injected momentum.

Figure~\ref{fig:snap_m2p} displays XY-slices of the M2P models at final stage
of evolution.  The morphology of the SN bubble is determined by the structure
of the ambient medium. Due to the short cooling time in model M2P-n10, the hot
gas persists only deep inside of the SN bubble; this is in contrast to model
M2P-n0.1, where the shell is still thin and the hot gas fills up the interior
of the SN bubble.  Figure~\ref{fig:tevol_multi} plots time evolution of (a) the
mean radius of the combined SNR $\rsnr$, (b) the effective radius of hot gas
volume $\rhot\equiv (3V_{\rm hot}/4\pi)^{1/3}$, (c) momentum $\psnr$, (d) hot
gas mass $\Mhot$, (e) thermal energy $E_{\rm th}$, and (f) kinetic energy
$E_{\rm kin}$.  Thick lines show M2P and thin lines E2P models, with mean
ambient density $n_0=0.1$ (blue), 1 (green), and 10 (red).  Note that
$\rsnr<\rhot$ at early stages when the SNR is filled with hot gas since the
mass-weighted radius is closer to the shell radius in the CNM, and
$\rsnr>\rhot$ at later stages when the hot gas is depleted (see right column in
Figure~\ref{fig:snap_m2p}). 

Evidently, multiple SNe are able to inject more momentum when the ambient
density is lower, similar to the case of single SNe.  However, the dependence
on ambient density is even weaker than in the single SN case.  The final
momenta for $n_0=0.1,$ 1, and 10 are $p_{\rm final}=2.2$, 1.7, and
$1.4\times10^6\momunit$, respectively.  
There is negligible increase in momentum shortly after the final SN event, for
all cases.
Comparing multiple SNe to a single SN with the same total energy, the ratios of
the total momentum from M2P models to E2P models are 61\%, 85\%, and 98\% for
$n_0=0.1$, 1, and 10, respectively.  The final momentum per SN event fitted to
these results is $1.7 \times 10^5 \momunit n_0^{-0.1}$.

The differences among the M2P models can be understood partly based on the
ratio of the SN time interval $\tsn$ to the shell formation time $\tsf$ for the
ambient medium.   The interval $\tsn =0.1 \Myr$ corresponds to $\tsf>\tsn$,
$\tsf\sim\tsn$, and $\tsf<\tsn$, for $n_0=0.1$, 1, and 10, respectively.  For
M2P-n10, new SN energy is injected only after completely exhausting the
previous SN,  since the interval is much longer than the shell formation time
$\tsf=0.012\Myr$ in the ambient medium.  The injection of momentum is thus in
discrete events, and the mass of hot gas never reaches values much larger than
that from a single SN event with $\ESN=10^{51}\erg$.  Model E2P-n10 has a much
higher peak $\Mhot$ than that of M2P-n10, because $\Msf$ is nearly linearly
proportional to $\ESN$ (see Equation (\ref{eq:Msf})).  However, $\Mhot$ drops
rapidly after the early shell formation for model E2P-n10.  In model M2P-n10,
hot gas is replenished after each new SN event, resulting in larger occupying
volume than E2P-n10. Both mass and volume of hot gas decrease after the last SN
due to depletion of hot gas by cooling, while the mean radius of the SNR keeps
increasing. 

For M2P-n0.1, the evolution is quite different from M2P-n10.  The mass of hot
gas reaches nearly an order of magnitude higher than for M2P-n10, and as the
hot medium does not fully cool in between events, the momentum injection
becomes almost continuous rather than discrete.   The peak in $\Mhot$ occurs at
nearly the same time for M2P-n0.1 as for E2P-n0.1, although for both the hot
gas is rapidly depleted after this time.  The volume occupied by hot gas keeps
increasing as $\rsnr$ for both models.

The evolution of model M2P-n1 is intermediate between that of M2P-n10 and
M2P-n0.1.  The mass of hot gas stays close to constant between the first and
last SN, with only slight dips.  The hot gas volumes for M2P-n1 and E2P-n1
converge with each other.  The momentum acquisition for model M2P-n1 is
therefore smoother than for model M2P-n10, and more discrete than for model 
M2P-n0.1.  

\section{SUMMARY \& DISCUSSION}\label{sec:sumndis}

Utilizing the hydrodynamic code \emph{Athena} with optically thin radiative
cooling, we simulate the expansion of radiative SNRs on three-dimensional
Cartesian grids.  For all models, each explosion is numerically initiated by
deposition of $E_{\rm SN}=10^{51}\erg$ in the form of thermal energy within a
volume of radius $\rinit$.  We consider three types of simulations: a single SN
explosion in a uniform medium (SU models; Section~\ref{sec:single}); a single
SN explosion in a two-phase medium (S2P models; Section~\ref{sec:single_2p});
and multiple SN explosions in a two-phase medium (M2P models;
Section~\ref{sec:multi_TI}).  We cover a range of mean density $n_0=0.1-100$.
For two-phase models, the cloud and intercloud densities are respectively an
order of magnitude above and below $n_0$.

With the SU simulations, we first delineate the detailed evolution of 
radiative SNRs, in comparison to analytic solutions and previous
one-dimensional (spherical) numerical simulations. Among other 
quantities, we evaluate the time, radius, mass, and momentum of the SNR at
shell formation, and the final momentum of the expanding cooled shell.  We then
conduct a wide, systematic parameter search to ascertain numerical conditions
for ``consistent convergence'' (convergence for all evolutionary stages) of SNR
evolution. We find that $\rinit/\rsf<1/3$, where $\rsf$ the size of SNR at
shell formation, guarantees consistent convergence, and the final momentum is
within $25\%$ or $18\%$ of high-resolution benchmarks provided that the grid
resolution is $\Delta/\rsf<1/3$ or  $\Delta/\rsf<1/10$, respectively. 

We next conduct similar studies using more realistic ambient conditions,
consisting of a two-phase ISM characterized by irregular CNM clouds embedded in
the WNM.  We show that the same numerical convergence conditions apply to the
two-phase medium as the uniform medium, and evolution follows a similar course,
with comparable hot gas mass and momentum injection.  Finally, we investigate
the momentum injection by multiple SNe in a two-phase medium, using the same
numerical criteria to set the initial SNR size.  The details of evolution
depend on the interval between SNe compared to the shell formation time (which
decreases with increasing mean ambient density), but the final momenta of the
expanding (combined) SNRs are similar in all cases.

Our main findings and implications are summarized below.

\begin{itemize}
\item[1.] \emph{Stages of SNR Evolution} --  

Our simulations confirm the well-known evolutionary stages of a SNR subsequent
to early free expansion \citep[e.g.,][]{1974ApJ...188..501C,
1988ApJ...334..252C,1998ApJ...500...95T}: an energy conserving Sedov-Taylor
(ST) phase, shell formation when the remnant becomes radiative, a
cooling-modified pressure-driven snowplow (PDS) phase, and a final
momentum-conserving expansion phase.  In a uniform medium, the shell formation
epoch is well defined by the time at which the mass of hot gas attains its
maximum, $\sim 10^3 \Msun$.  Table~\ref{tbl:uniform} shows that the analytic
estimate for $\tsf$ from  Equation (\ref{eq:tsf}) provides a reasonable match
with our numerically-measured $\tsfnum$; this also agrees with previous
numerical findings for shell formation based on spherical one-dimensional
simulations \citep{1998ApJ...500..342B,1998ApJ...500...95T}.  The radius, total
SNR mass, and outward radial momentum at shell formation are also close to
analytic estimates.

In classical theory \citep[e.g.,][]{1988RvMP...60....1O,2011piim.book.....D},
an adiabatic PDS stage (with constant interior mass and
$\Phot\propto\rsnr^{-5}$) follows shell formation, but this ideal situation is
not realized in practice \citep[see
also][]{1988ApJ...334..252C,1998ApJ...500...95T}. Instead, the hot interior
expands into the back of the shell, where it condenses and cools.  The mass of
hot gas and interior pressure drop by an order of magnitude within a few times
$\tsf$. With rapidly-falling interior pressure, the momentum added to the shell
during the PDS stage is modest (see below).  Figure~\ref{fig:tevol_norm} shows
that for a uniform medium, SNR evolution through ST, shell formation,
cooling-modified PDS, and momentum-conserving stages is essentially congruent
for varying ambient density, when normalized by the SNR properties at
$\tsfnum$. 

For a two-phase ambient medium, Figure~\ref{fig:tevol_2p} shows similar
evolutionary stages to the case of a uniform medium.  Although energy losses
commence earlier (since dense clouds radiate after they are shock-compressed),
the mass of hot gas reaches a maximum (and then rapidly declines) only when a
strongly-cooling shell forms in the low-density volume-filling intercloud
medium. The numerically-computed shell formation time is in between the values
of  $\tsf$ computed using the mean ambient density and using the density of the
intercloud medium (Table \ref{tbl:2p}).  A hot, low-density and low pressure
bubble is left behind as dense cloudlets and an irregular shell of
swept-up/cooled intercloud gas expand outward.

When multiple explosions occur sequentially in the same location, the evolution
is either highly impulsive or relatively smooth depending on the interval
between SNe compared to the shell formation time in the ambient medium (see
Figure \ref{fig:tevol_multi}).  Similar to the case of single SNe, however,
cooling of the interior implies that a constant asymptotic momentum is reached
shortly after the last SN event.   

Most of our simulations consider evolution of SNR in an unmagnetized medium.
However, in the Appendix we include results for the evolution of magnetized models
with a range of plasma beta parameter for the background medium ($\beta\equiv
P_{\rm ISM}/P_{\rm mag}$ in the range 0.1 to 10). For $\beta=1$ and 10,
evolution is essentially indistinguishable from the unmagnetized case up to
about $10\tsf$.  For the $\beta=0.1$ case, at late stages the shell thickens
and becomes anisotropic due to magnetic pressure and tension forces, and 
at very late stages the SNR becomes elongated along the magnetic field
\citep[e.g.,][]{2006ApJ...641..905H}.

\item[2.] \emph{Momentum Injection by a Single SN} --

Our simulations show that most of the momentum of the expanding remnant is
acquired during the ST stage of evolution.  For uniform ambient conditions, the
final momentum is only $50\%$ larger than the momentum $\psf$ at shell
formation, at all values of the ambient density (see
Figure~\ref{fig:tevol_norm}). With the very weak dependence of $\psf$ on the
ambient medium density (see Table~\ref{tbl:uniform}) as predicted by Equation
(\ref{eq:psf}), the final momentum injected by a single SN in a uniform medium
varies $\propto n_0^{-0.16}$, yielding $\sim (1-4)\times10^5\momunit$ for
$n_0=100 - 0.1$ (see Equation (\ref{eq:pfinal})).

We show in Appendix that the momentum injection is the same for a magnetized
medium as for the unmagnetized case, when the plasma beta parameter exceeds
0.1. This is because most of the momentum is acquired during stages when
thermal pressure within the SNR greatly exceeds magnetic forces.  

For a two-phase ambient medium, most of the momentum is acquired during the 
period prior to shell formation in the low-density intercloud medium (Figure
\ref{fig:tevol_2p}).  For given mean density, there is some variation in when
and how much total momentum is gained depending on the locations and sizes of
dense clouds around the explosion site (see Figure~\ref{fig:rnorm_2p}), at a
level of a few tens of percent.  Averaging over different realizations,
however, the momentum at the shell formation time (when the hot gas mass is
maximal) is within 20\% of the value predicted by Equation (\ref{eq:psf}) using
the mean density averaged over both dense clouds and low-density intercloud
medium (see ``WHOLE'' rows in Table~\ref{tbl:2p}). After shell formation, the
momentum increases by at most 50\%.  For a cloudy medium, the final momentum
from a single SN varies $\propto n_0^{-0.17}$ is also in the range
$(1-4)\times10^5 \momunit$ when the $n_0=100 - 0.1$.  (see Equation
(\ref{eq:pfinal_2p})).

Our simulations show that the final momentum injected by a single SN is an
order of magnitude larger than the initial momentum of the ejecta.  This
conclusion is insensitive to the mean density of the ambient medium, and
irrespective of strong inhomogeneity in the ISM surrounding the explosion site.
As a simple quantitative mnemonic, the momentum injected by a single SN is
comparable to the initial energy of the blast ($10^{51}\erg$) divided by the
speed of the expanding shock wave when it becomes radiative ($\approx
200\kms$).

\item[3.] \emph{Implications for SN Momentum Injection in the ISM} --

For scales larger than an individual star forming region, the amount of
momentum injection to the ISM by a SN dominates other possible sources of
feedback associated with earlier stages of a massive star's evolution, such as
stellar winds, radiation pressure, and \ion{H}{2} region expansion \citep[see
e.g.,][for analytic estimates of inputs from these
sources]{2002ApJ...566..302M, 2010ApJ...709..191M,2011ApJ...731...41O}.  In
recent work, a simple momentum injection model (at a level $p_{\rm
final}=3\times10^5\momunit$ comparable to that found here) has been adopted to
treat SN feedback, driving turbulence in the ISM.  Simulations implementing
this ``pure momentum'' approach have been used to model molecule-dominated
starburst regions \citep{2012ApJ...754....2S}, multiphase atomic-dominated
regions \citep{2011ApJ...743...25K,2013ApJ...776....1K}, and barred spiral
galaxies \citep{2013ApJ...769..100S,2014ApJ...792...47S}.  Among the successes
of these models are reaching realistic turbulence levels in the diffuse ISM,
and achieving states of self-regulated star formation with rates that are
consistent with observations  \citep[see
also][]{2010ApJ...721..975O,2011ApJ...731...41O}.  In addition,
\citet{2014ApJ...786...64K} demonstrated that synthetic \ion{H}{1} 21 cm lines
reconstructed from the models of \citet{2013ApJ...776....1K} are consistent
with the characteristics of real 21 cm line observations by \citet[][see also
\citealt{2014ApJ...793..132S}]{2013MNRAS.436.2352R}.

In many recent investigations studying individual galaxies
\citep[e.g.,][]{2012MNRAS.421.3488H,2013MNRAS.429.3068T,2013ApJ...770...25A} or
galaxy formation in a cosmological context
\citep[e.g.,][]{2011ApJ...742...76G,2013MNRAS.428..129S,2014MNRAS.445..581H},
it has been found that models with only SN feedback  are unable to limit star
formation rates to realistic values.  Concerns about the inefficacy of SNe has
prompted increased attention to other modes of feedback, a valuable development
in itself (especially for understanding the evolution of molecular clouds at
stages prior to the first SN).  However, these concerns are to some extent
misplaced, and reflect numerical rather than physical limitations.  In
particular, it can be difficult to capture the crucial (but spatially small
scale) ST stage of evolution -- when most of the momentum is imparted -- in
simulations that must follow extremely large spatial scales and cannot afford
high resolution for the ISM.  In item 4 below,  we summarize the numerical
requirements for properly resolving SN momentum injection.

One reason for the current uncertainty about feedback from SN is that the real
ISM is far from homogeneous, whereas previous focused studies of SNR expansion
have mostly considered homogeneous (or at least isotropic) environments.  For
example, \citet{2013ApJ...770...25A} applied only the initial ejecta momentum
$\pfr$ rather than the momentum at the end of ST phase $\psf$
or the final momentum in the momentum conserving phase $p_{\rm final}$,
citing concerns regarding inhomogeneity of the site of SN explosion.  However,
as seen in Section~\ref{sec:single_2p}, inhomogeneity of the background medium
does not significantly affect the total momentum injection compared to the
uniform-medium case. The effective spatial scale for momentum deposition is
larger in an inhomogeneous medium than in a uniform medium, however, because
the shock propagates further in low-density intercloud gas before it becomes
radiative.  

Several independent studies contemporaneous with our own have also studied
momentum injection by SNR in the inhomogeneous ISM.  Although different groups
have considered different environments and physical ingredients in their
simulations, all have reached similar conclusions: the final injected momentum
in a homogeneous and inhomogeneous medium are comparable to each other, and in
good agreement with the results of our simulations.  This result is
irrespective of initialization methods to impose inhomogeneity; we begin with a
two-phase cloud/intercloud medium produced by thermal instability, while
\citet{2014arXiv1409.4425M} adopt a lognormal density distribution,
\citet{2014arXiv1410.0011W} adopt a fractal density structure, and
\citet{2014arXiv1410.7972I} initialize based on a turbulent velocity field.
The latter two studies also consider SNe within radially stratified clouds.
\citet{2014arXiv1410.0011W} and \citet{2014arXiv1412.0484G} find that explosion
of the SN within a pre-existing low density \ion{H}{2} region can enhance the
momentum up to $\sim 5\times 10^5\momunit$
(including the momentum injected by \ion{H}{2} region expansion) for solar
metallicity. 

\item[4.] \emph{Numerical Conditions for Correct SNR Evolution} --

At the resolution of current galaxy formation simulations (tens of parsecs or
larger), and with realistic ISM densities, the ST phase cannot be properly
resolved.  This leads to the oft-cited ``overcooling problem,'' namely that the
energy of a SN is radiated away without having  much impact on the ISM, if it
is initially deposited in too large an initial volume or mass.  Although this
problem is well known, there has not previously been a systematic study to
determine the numerical requirements needed to avoid it.  

Here, we have conducted a large set of simulations varying both the grid
resolution ($\Delta$) and the initial size of the SNR ($\rinit$).  We find that
$\rinit/\rsf<1/3$ is necessary to follow the hot gas evolution properly.  We
also find that the final momentum is within 18\% and 25\% of high-resolution
benchmarks provided $\Delta<\rsf/10$ and $\rsf/3$, respectively. If the SN
explosion occurs within an ambient medium density of $n_0=1$ or 10,
$\rsf=23\pc$ or $9\pc$ thus gives a requirement of $\Delta\sim\rinit<7\pc$ or
$3\pc$, respectively, for $\Delta=\rinit=\rsf/3$  

For practical purposes in an inhomogeneous medium, the resolution requirements
can be expressed in terms of a maximum enclosed mass within the initial SNR.
The criterion $\rinit/\rsf<1/3$, equivalent to $\Minit<\Msf/27$, corresponds to
an enclosed mass of $\Minit<(20-110)\Msun$ for $n_0=100$-0.1.  Interestingly,
this is similar to the condition adopted in \citet{2006ApJ...653.1266J}, who
argued that $\Minit\sim60\Msun$ would yield an initial temperature in the range
of $10^6$-$10^7\Kel$, for which the cooling rate would not be excessive.
\citet{2006ApJ...653.1266J} and subsequent studies by their colleagues
\citep{2009ApJ...704..137J,2012ApJ...750..104H} modelled local galactic disks
with large vertical domains, including hot gas created by SN shocks.  The
numerical resolution of $1\pc$ combined with the condition $\Minit\sim60\Msun$
and dispersed SN events allowed them to correctly resolve the evolution of most
SNRs.  However, these models did not include self-gravity, which can lead to
higher densities at the sites of SN events such that $1\pc$ grid resolution
would be marginal or insufficient (considering that $\rsf/3\sim 1\pc$ for
$n_0=100$); alternative approaches are discussed below.

\item[5.] \emph{Alternative Feedback Prescriptions}

If ambient densities are not too high, and the grid is sufficiently fine, the
needed resolution requirements $\rinit/\rsf<1/3$ (or $\Minit<\Msf/27$) and
$\Delta/\rsf < 1/10$ can be met, and both the hot gas evolution and momentum
injection of a SN can be properly followed by injecting thermal energy and
allowing the SNR to progress through ST and radiative stages.  We recommend
this approach for best fidelity in modeling SN feedback.    However, this is
not always possible in numerical simulations, even with adaptive mesh
refinement techniques.  

In situations where it is impossible or impractical to resolve SNR evolution
including the ST stage, an alternative is to inject momentum directly to the
gas surrounding the feedback site.  This will miss the direct effects of hot
gas in the ISM (including potential losses of this gas as a galactic wind), but
it will still drive turbulence in the warm and cold phases of the ISM, which
enables star formation to be correctly self-regulated
\citep{2012ApJ...754....2S,2011ApJ...743...25K,2013ApJ...776....1K}.  To allow
for the (modest) dependence of momentum injection on ambient conditions, we
recommend that Equation (\ref{eq:pfinal_2p}) be used to set the value of the
momentum after the mean density in the numerical feedback region is measured.

In a spirit similar to the approach that we recommend, resolution-conditioned
approaches to feedback in galaxy formation simulations have been implemented
recently by  \citet{2014MNRAS.445..581H} and \citet{2014ApJ...788..121K}.  In
both studies, SN feedback is assigned either by the ejecta directly to nearby
grid zones or SPH particles if the local resolution was high, or as ``final''
(post-ST) momentum otherwise.  In their SPH simulations,
\citet{2014MNRAS.445..581H} compared the smoothing length with the shell
formation radius $\rsf$ ($R_{\rm cool}$ in their nomenclature) derived by
\citet{1988ApJ...334..252C} to determine which prescription to apply.  In their
AMR simulations, \citet{2014ApJ...788..121K} apply either the initial ejecta
momentum or a ``final'' (post-ST) momentum to zones depending on whether the
mass in a given feedback sector is greater or less than a SNR would have at the
``transition'' time of \citet{1998ApJ...500..342B}.  The resolution
requirements we recommend, combined with our calibrations of momentum injection
for varying mean density, may be useful in refining and further developing
conditional-feedback prescriptions similar to these.  The resolution
requirements $\rinit/\rsf<1/3$ and $\Delta/\rsf < 1/10$ that we have obtained
through convergence studies could be immediately applied in grid-based
simulations that use similar finite-volume methods to those of the {\it Athena}
code.  For SPH simulations, it would be straightforward to conduct an analogous
study to determine numerical requirements for convergence in the  hot gas mass
evolution and net momentum injection.

\item[6.] \emph{Multiple SNe and Superbubbles/Galactic Winds} --

When a sufficiently massive star cluster forms, one can expect not a single but
multiple SN explosions to impact the surrounding ISM.  Many previous studies of
multiple SNe have focused on what requirements must be met for a superbubble to
break out of the disk and drive a wind
\citep[e.g.,][]{1986PASJ...38..697T,1987ApJ...317..190M,1989ApJ...337..141M,1990A&A...237..207T,1992ApJ...388...93K,1992ApJ...388..103K,2013MNRAS.434.3572R}.
In Section~\ref{sec:multi_TI}, we instead quantify momentum injection to the
ISM, which is key to settling the turbulence level and rate of star formation.
In this paper, we consider only a highly-idealized model of the impact of a
moderate-sized star cluster on the surrounding ISM, in which SN blasts are
intermittent.  We explode 10 SNe with time interval of $\tsn=0.1\Myr$ (M2P
models) in a two-phase medium with density of $n_0=0.1$, 1, and 10, following
evolution for a total of $2\Myr$ after the first explosion (and $1\Myr$ after
the last explosion).  
These specific conditions limit the SN bubble size smaller than or comparable
to the disk scale height, allowing us to explore the momentum injection by
a superbubble in a uniform medium.

When $\tsf<\tsn$, as for model M2P-n10, momentum is
injected by individual events in a discrete fashion, whereas in the opposite
limit of $\tsf > \tsn$ for model M2P-n0.1, momentum injection is almost
continuous.  In spite of these differences, there is only $\sim 50\%$ variation
in the final momentum measured in the three M2P models ($p_{\rm final} =
14-22\times 10^5 \momunit$).  As for single SNe, the final momentum per SN
event in the M2P models is comparable to the prediction $\psf$ of Equation
(\ref{eq:psf}).  For comparison, we also conduct comparison models with a
single SN having $\ESN=10^{52}\erg$ (E2P models); the final momenta for E2P
models are a factor $1 - 2$ larger than for the corresponding M2P models.

In recent one-dimensional simulations of sequential SNe,
\citet{2014MNRAS.443.3463S} have found that with a sufficiently large number of
SNe (i.e. a sufficiently small mean interval $\tsn$), the evolution of the SN
bubble approaches the limit of continuous energy injection and an adiabatic
interior, as in \citet{1977ApJ...218..377W} \citep[see
also][]{1981Ap&SS..78..273T,1987ApJ...317..190M,1988RvMP...60....1O,1992ApJ...388...93K,1992ApJ...388..103K}.
For continuous energy injection, the radius evolves with $\eta=3/5$ rather than
$\eta=2/5$ or $1/4$.  Substantial energy (about 10-30\% of total explosion
energy) can remain in this limit \citep[see also][]{2015MNRAS.446.1703V}, as is
required to explain observations of systems such as M82
\citep[e.g.,][]{2009ApJ...697.2030S}. 

Although our M2P models create large, low-density hot volumes that resemble
observed superbubbles, here we have not considered a stratified medium or
highly-dynamic initial conditions for the ambient warm/cold ISM.  
For a realistic cluster lifetime with $\tsn=0.1\Myr$, however, the total
number of massive stars that could undergo core collapse is closer to 300,
which would drive a superbubble expansion over a much longer period, such that
its radius could easily exceed the disk scale height.  More massive clusters
would have even shorter mean intervals between SN events, and could easily lead
to superbubble breakout. For more comprehensive studies for superbubble
evolution, stratification of the disk should thus be taken into account,
in addition to the small-scale inhomogeneity that we have included.
A natural extension of the present work is to include both effects, together
with multiple SNe, to study the superbubble breakout and creation of galactic
winds that may occur when a very massive cluster forms.  

\end{itemize}

Finally, we note some caveats related to limitations of the models we have
presented.  Because our resolution is insufficient to capture ablation of small
clouds by the instabilities (Rayleigh-Taylor and Kelvin-Helmholtz) arising from
shock-cloud interactions \citep[e.g.,][]{1994ApJ...420..213K}, our models may
underestimate cooling within the SNR.  Similarly, as we do not include thermal
conduction
\citep[e.g.,][]{1977ApJ...211..135C,1981ApJ...247..908C,1982ApJ...252..529B},
we may be missing an increase in the mass of hot gas and enhanced cooling due
to evaporation of and radiation from clouds.  Compared to SNR evolution within
a uniform intercloud medium, cloud ablation and evaporation lead to earlier
shell formation as the effective density of the ambient medium is higher than
that of the intercloud medium.  The momentum injection could also be reduced
somewhat due to additional radiative losses of energy.  Although our simulation
cannot fully capture cloud ablation and evaporation, these effects have been
captured to some extent: in models S2P, the shell formation radius is smaller
than that in a purely warm medium (see Table~\ref{tbl:2p}).  Also, the final
injected momentum is smaller in models S2P than SU (see Equations
(\ref{eq:pfinal}) and (\ref{eq:pfinal_2p})), and among 10 different
realizations a slightly smaller momentum is obtained when the SNR sweeps up
more cold clouds.

However, it is also important to note that effects of thermal conduction and
cloud ablation become more complex for a magnetized medium, as is true for the
ISM. Thermal conduction is anisotropic, so that the heat flux across the
magnetic field is largely suppressed \citep{1962pfig.book.....S}.  When a shock
encounters a cloud, the magnetic field becomes wrapped around the cloud (piled
up in front of the cloud, and trailing downstream).  This geometry limits heat
flux from the hot medium to the cloud.  In addition, hydrodynamic instabilities
are strongly suppressed with near-equipartition magnetic fields $\beta\sim1$
\citep[e.g.,][]{1994ApJ...433..757M,2005ApJ...619..327F,2008ApJ...680..336S}.
Calculations of shock-cloud interactions including magnetic fields, anisotropic
thermal conduction, and radiative cooling by \citet{2008ApJ...678..274O} have
shown that magnetic fields with moderate strength ($\beta=4$) indeed suppress
heat transfer and hydrodynamic instabilities \citep[see
also][]{2013ApJ...766...45J}.  In future work, it will be interesting to extend
the SNR expansion simulations of this paper to include magnetic fields and
anisotropic conduction, in order to quantify possible changes in the momentum
injection, maximum mass of hot gas, and other basic properties.  In addition,
complete models would include cosmic ray pressure, which can potentially boost
the late-time SNR expansion and momentum.  The magnitude of this effect depends
on the diffusivity of the cosmic ray fluid, which is becoming increasing well
characterized from recent hybrid kinetic simulations
\citep[e.g.,][]{2014ApJ...794...47C}.

\acknowledgements

We are grateful to the referee for an extremely thorough report, which 
helped us to improve the manuscript.
This work was supported by Grant No. AST-1312006 from the National
Science Foundation.  Part of this project was conducted during a visit
to the KITP at U.C. Santa Barbara, which is supported by the National
Science Foundation under Grant No. NSF PHY11-25915.
This research made use of {\tt YT} visualization and data analysis package
\citep{2011ApJS..192....9T}.

\appendix

\section{Single SN in Uniform, Magnetized Medium}

\begin{figure}
\epsscale{1.0}
\plotone{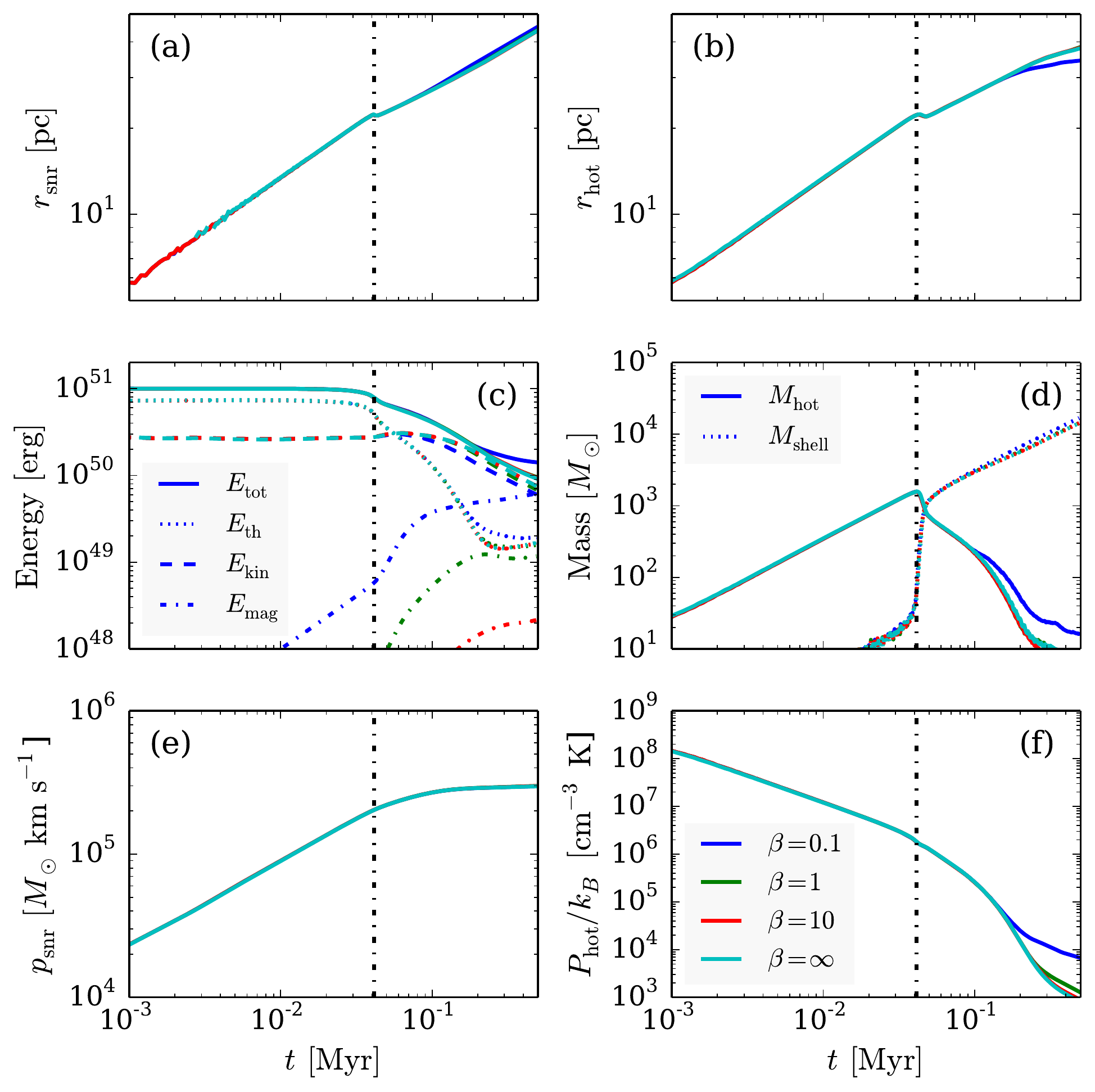}
\caption{
Time evolution of models analogous to SU-n1 ($n_0=1\pcc$ and
$\ESN=10^{51}\erg$), but with initial uniform magnetic fields characterized by
varying $\beta$.  (a) mass-weighted radius, (b) the effective radius of the hot
gas volume, (c) total, thermal and kinetic energies, (d) mass of interior hot
gas and shell, (e) total radial momentum, and (f) pressure of interior hot gas.
Grid resolution is $\Delta = 1/2\pc$ and the initial SNR radius is $\rinit =
3\pc$.  The vertical dot-dashed lines in each panel denote the predicted shell
formation time $\tsf=4.4\times10^4\yr$ (Equation (\ref{eq:tsf})) for this model
without magnetic fields. 
\label{fig:tevol_mhd}}
\end{figure}

\begin{figure}
\epsscale{1.0}
\plotone{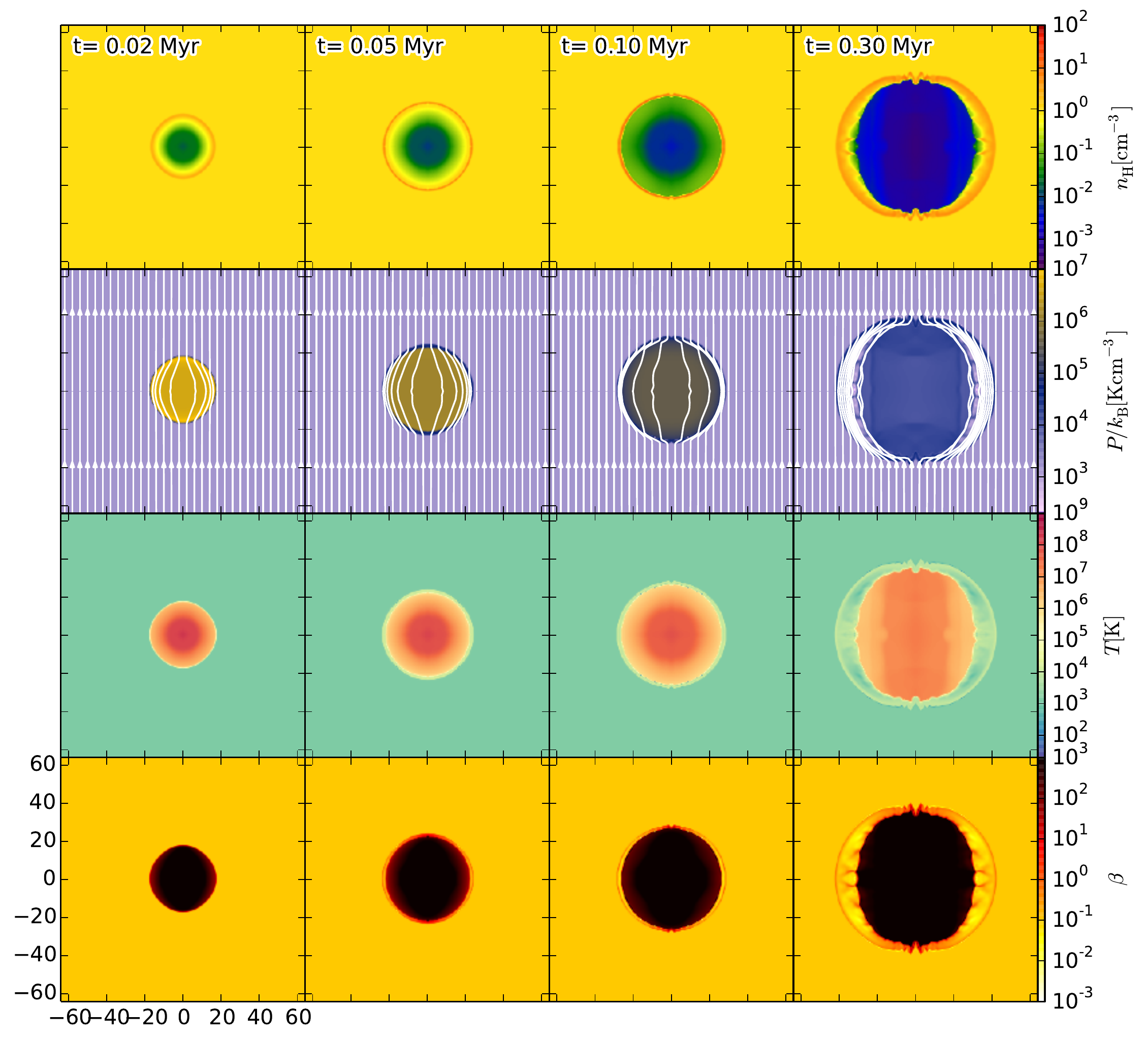}
\caption{
YZ-slices from magnetized model with $n_0=1$ and $\beta=0.1$, corresponding to
$B_{z,0}=7.2\mu G$. From top to bottom, logarithmic color scales show number
density, pressure, temperature, and plasma beta. From left to right, columns
correspond to snapshots at $t/\tsf\sim1/2$, 1, 2, and 7.
Magnetic field lines are shown in the second row (pressure).
  \label{fig:snap_mhd}}
\end{figure}

The effect of magnetic fields in SNR evolution was initially  investigated
using spherically symmetric one-dimensional simulations, in which magnetic
pressure forces are included but the full MHD equations are not solved
\citep{1974ApJ...188..501C,1992ApJ...392..131S} (this can approximately
represent an equatorial band tangent to a uniform external magnetic field).
The strong magnetic pressure within the compressed shell results in a thicker
shell and recompresses hot gas in the later times.  \citet{2006ApJ...641..905H}
followed the long-term evolution of a SNR in a magnetized medium using
axisymmetric two-dimensional simulations, and showed that the magnetic tension
force also helps broaden the shell and compress the hot gas \citep[see
also][for supperbubble expansion]{1991ApJ...375..239F}. Magnetic fields
play an important role in determining the volume of the hot gas in very late
stages of a SNR, and therefore affects the porosity of the ISM.

Here, we are interested in assessing potential effects of magnetic fields on
momentum injection by SNRs in the ISM.  We have re-run the set of SU models,
but now include a uniform magnetic field in the $z$-direction
$\mathbf{B}_0=(0,0,B_{z,0})$.  The magnetic field strength is set by using the
plasma beta 
\begin{equation}\label{eq:beta}
\beta\equiv \frac{8\pi P_{\rm ISM}}{B_{z,0}^2},
\end{equation}
where the thermal pressure of the ISM in our models varies with 
the mean density, $n_0\equiv n_H/1\pcc$, as $P_{\rm ISM}=1500 n_0\kbol\Punit$.
For reference, the magnetic field strength is
\begin{equation}\label{eq:Bz}
B_{z,0}=3.2\mu G \rbrackets{\frac{P_{\rm
ISM}/\kbol}{3000\Punit}}^{1/2}{\beta}^{-1/2} = 2.3(n_0/\beta)^{1/2}\;\mu G .
\end{equation}

Figure~\ref{fig:tevol_mhd} plots time evolution of physical properties of the
SNR in models SU-n1 with $\beta=0.1$, 1, 10, and $\infty$. All models follow
almost the same evolution except at late time in the model with $\beta=0.1$. It
is obvious that the magnetic energy in the shell becomes comparable to the
thermal energy of the shell only for $\beta=0.1$ after shell formation (see
Figure~\ref{fig:tevol_mhd}(c)).  As reported in earlier work, a strong
magnetic field within the shell recompresses the interior hot gas (smaller
$\rhot$ at late time for $\beta=0.1$),
leading to larger $\Mhot$ and $\Phot$ at late times (see
Figure~\ref{fig:tevol_mhd}(d) and (f)). 
However, the radial momentum is almost the same for all cases.
This is because the majority of the momentum is injected to the ISM before the
magnetic energy in the shell is very large.  Even for the $\beta=0.1$ case, full
momentum is acquired by (2-3)$\tsf$, when the magnetic energy just begins to
dominate the thermal energy.

Figure~\ref{fig:snap_mhd} displays the structure of the model with $n_0=1$ and
$\beta=0.1$ in the YZ plane.  Early evolution is the same as for the
unmagnetized model (left three columns, see Figure~\ref{fig:snap}), while
non-spherical structure emerges in the latest stage (rightmost column). The
outer shock can propagate farther in the direction perpendicular to the
magnetic fields (fast mode), while shell broadening results in compression of
the interior hot gas.

We have also performed simulations analogous to models SU-n0.1, SU-n10, and
SU-n100 with $\beta=0.1$, 1, and 10. We find that the normalized evolution for
all cases is ``congruent'' (as in unmagnetized models) up to 2 or 3 $\tsf$, so
that the final momentum remains unchanged. This suggests that ISM magnetic
fields with reasonable strengths of $\beta=0.1-10$ may be able to alter late
time evolution and porosity of the hot gas, but would not significantly affect
the injected momentum \citep[see also][]{2014arXiv1410.7972I}.

\end{document}